\documentclass[11pt]{article}
\usepackage[T1]{fontenc}
\usepackage{amsmath,amssymb,amsthm}
\usepackage{authblk}
\usepackage{graphicx}
\usepackage{float}
\usepackage{algorithm,algpseudocode}
\usepackage{placeins}
\usepackage{booktabs,microtype,balance,xcolor, comment} 
\usepackage[round,authoryear]{natbib}
\usepackage[hidelinks]{hyperref}
\usepackage{xurl}
\usepackage[all]{hypcap}
\newtheorem{theorem}{Theorem}
\newtheorem{lemma}{Lemma}
\newtheorem{proposition}{Proposition}
\newcommand{\E}{\mathbb{E}}
\newcommand{\Var}{\operatorname{Var}}
\newcommand{\Cov}{\operatorname{Cov}}
\newcommand{\CF}{\mathrm{CF}}

\title{Bias correcting transaction cost models for characteristic function based estimators}%
\author[1]{Abhik Rana \thanks{abhik.rana@ashoka.edu.in}}
\author[1]{Sarvesh Ravichandran Iyer \thanks{sarvesh.iyer@ashoka.edu.in}}
\author[1]{Sandeep Juneja \thanks{sandeep.juneja@ashoka.edu.in}}

\affil[1]{Ashoka University}
\date{}

\begin{document}

\maketitle

\begin{abstract}
Effective transaction cost is a necessary step for devising any competing financial trading strategy in the absence of quote data. Classically, Roll's model estimates the transaction cost using the first order autocovariance of the difference between daily transaction prices, while Hasbrouck’s Bayesian approach uses iterative posterior simulation. However, Roll's model fails when the first order autocovariance becomes positive, while Hasbrouck's estimator is computationally intensive, especially when many securities are involved. We develop empirical characteristic function methods with a first order bias correction for spread estimation, and establish the asymptotic distribution of the corrected estimator under suitable conditions. Our numerical experiments illustrate the bias–variance tradeoff of characteristic-function methods compared to Roll's model, and their computational advantage over Hasbrouck's model.
\end{abstract}

\section{Introduction}\label{sec:intro}
During financial portfolio optimization, accounting for transaction costs  is a key step in improving existing trading strategies (see \cite{magill1976portfolio,constantinides1979multiperiod,constantinides1986capital}). Estimates of these costs help us accurately assess trading strategies and study the relation between liquidity and returns \citep{hasbrouck2009}.

In the absence of quotes and trade directions, one typically infers transaction costs from their prices, by separating changes in the theoretical efficient price from changes caused by trades across the spread. There are a number of reasons why one does this, ranging from measurement problems for bid-ask data \citep{roll1984} to their cumbersome usage, high cost of purchase, or complete absence; see \cite{hasbrouck2009, corwinschultz2012}.

A foundational model in transaction cost estimation is that of \cite{roll1984}. Essentially, the model assumes that under perfect information, trading would take place at some "efficient" price which captures all market and trading information up till that point. In reality, intermediaries such as dealers and exchanges must be compensated for effectuating transactions, and as information arrives, buyers and sellers constantly update their prices in a symmetric but stochastic fashion around the efficient price.

To be precise, at a time point $t \in \{0,\ldots,n\}$, the efficient market price $M_t$ and the trade price $P_t$ of a stock evolve according to the equations \begin{align}
M_t =& M_{t-1}+U_t, \text{ and}    \label{roll1}\\
P_t =& M_t + c \,Q_t.    \label{roll2}
\end{align}

Here, $U_t \sim N(0,V)$ are i.i.d. normal deviations with some unknown \emph{innovation} variance $V>0$ indicating market trends, $c>0$ is the per-trade cost borne by both seller and buyer which we wish to estimate (also called the \emph{half-spread}), and $Q_t$ are i.i.d. random variates indicating the trade direction, with $Q_t = \pm 1$ with probability $\frac 12$ each. The collections $\{U_t\}_{t =1,\ldots,n}$ and $ \{Q_t\}_{t=0,\ldots,n}$ are also independent of each other. 

If $Y_t = P_t - P_{t-1}$ denotes the price change process, then an easy calculation (see \cite{roll1984}) shows that $c = \sqrt{-Cov(Y_t, Y_{t-1})}$, which can therefore be estimated from data consisting of \emph{only} trade prices. Observe that for each $t=1,\ldots,n$,
\begin{equation}\label{yt}
Y_t = c(Q_t - Q_{t-1}) + U_t.
\end{equation}

In practice, however, it has been seen that $Cov(Y_t, Y_{t-1})$ can be positive for many values of $t$, leading to the estimate above being undefined \citep{choi1988,stoll1989,harris1990,george1991}. A number of alternatives have since been proposed, such as extended Roll's models \citep{glosten1987components, glosten1988estimating,huang1997components,lin1995trade}, the Corwin-Schultz estimator \citep{corwinschultz2012}, and the Abdi-Ranaldo estimator \citep{abdiranaldo2017}.

A popular model among these is that of \cite{hasbrouck2009}, who replaced the simpler equation \eqref{yt} with one of the form $$
Y_t = c (Q_t - Q_{t-1}) + \beta_m r_{mt}+U_t.
$$
Here, $r_{mt}$ is the return of the market asset at time $t$, and $\beta_m$ is a parameter that measures how much the asset return affects the change in trade prices. It is assumed that $r_{mt}$ is independent of $Q_t$. To address the issue of inadmissibility, a Bayesian approach is used to estimate the parameters $c, \beta_{m}$ and $V$. Namely, one assumes a Gaussian prior on $c$ and $\beta_m$, and an inverse-Gamma prior on $V$. The parameters of these priors are updated using one-step posterior estimation from the data. We refer the reader to \cite[Section C]{hasbrouck2009} for the complete details.

The Hasbrouck model offers a number of advantages, including better correlation with the actual market, thus affording wider applicability to a number of trading strategies \citep{hasbrouck2009, goyenko2009liquidity,hasbrouck2004liquidity}. However, the most significant drawback of the model is the intensive compute associated to Bayesian updation \citep{A, corwinschultz2012, goyenko2009liquidity}, along with poor performance under extremely high asset volatility \citep{jahanparvar2023when, mamun2026bidask}.

 Furthermore, each of the other aforementioned estimators also have inadequacies that have been reported in the literature. These discussions primarily have to do with \emph{bias}. For instance, estimation biases \citep{TRI}, specifically downward estimation biases \citep{A,nieto2018bid} have been recorded for popular estimators including Roll's and Hasbrouck's models. While our primary focus is on bias correction, other issues such as performance in illiquid markets \citep{TRI}, volatile environments \citep{johann2017best}, and low volume markets \citep{schestag2016measuring} have also been pointed out.

In this article, we build a characteristic function (CF) estimator of $c$ that targets bias reduction specifically, but also runs on minimal data and compute. We derive the joint bias and covariance expansion for a preliminary CF estimator, and tune it in a manner that reduces the leading order of the bias. The same expansion gives a criterion for a convex aggregate to reduce both leading absolute bias and mean square error (MSE) relative to Roll's model. Note that while earlier work has developed CF estimators in extended Roll models \citep{chenetal}, finite-sample bias remains a concern: they identify a downward distortion toward zero for small spreads in short samples and mitigate it by excluding problematic frequency-grid points. Their consistency and asymptotic normality results do not provide an explicit correction of the leading $n^{-1}$ bias term. We focus explicitly on this term in our analysis.

We then introduce independent pilots to select the frequency and aggregation weights in the estimator. Finally, we establish a confidence set which controls the probability of falsely certifying a gain.

The outline of this article follows. In Section~\ref{sec:model} we define a preliminary estimator. Section~\ref{sec:bias} gives the joint central limit theorem, bias-neutral frequency choice, and boundary limits comparing our estimator with Roll's estimator. Section~\ref{sec:aggregation} contains our main result: an aggregated estimator that reduces absolute bias and squared error relative to Roll's model.
Section~\ref{sec:selection} describes the pilot procedure. Finally, 
Sections~\ref{sec:simulations} and ~\ref{sec:discussion} contain the experiments and conclusion, respectively.

\section{Model and identification}\label{sec:model}
Recall Roll's model \eqref{yt}, with $c>0$ the half spread and $V>0$ the innovation variance. Define characteristic function of $Y_t$ at u to be $\phi(u) = \E[e^{iuY_t}]$.

We now define our estimator for $c$. For a frequency $a>0$, define
\begin{equation}\label{eq:phid}
\widehat\phi(a)=\frac1n\sum_{t=1}^n\cos(aY_t),
\ \widehat d(a)=1-\frac{\sqrt{\widehat\phi(2a)}}{\widehat\phi(a)^2}.
\end{equation}
Since $Y_t$ is symmetric, $\widehat\phi(a)$ estimates its characteristic function at $a$. Independence in Roll's model gives
\[
\phi(u)=\E\{\cos(uY_t)\}=e^{-Vu^2/2}\cos^2(cu),\quad u\in\mathbb R.
\]
The following identity eliminates the innovation variance.

\begin{lemma}\label{lem:identification}
If $0<ac<\pi/4$, then
\begin{equation}\label{eq:identity}
1-\frac{\sqrt{\phi(2a)}}{\phi(a)^2}=\tan^4(ac).
\end{equation}
\end{lemma}
The proof is in Appendix~\ref{app:identification}, and is a straightforward calculation. Taking the nonnegative fourth root and the principal inverse tangent suggests the estimator
\begin{equation}\label{eq:est}
\widehat c_{\CF}(a)=\frac1a\arctan\!\bigl(\widehat d(a)^{1/4}\bigr).
\end{equation}
Observe that $\widehat c_{\CF}$ is admissible when $\widehat\phi(a)>0$ and
$0\leq\widehat\phi(2a)\leq\widehat\phi(a)^4$. Our main result of this section is that inadmissibility occurs with an exponentially vanishing probability. Write
\begin{equation}\label{eq:xy}
x=\phi(a),\qquad y=\phi(2a).
\end{equation}
\begin{samepage}
Fix $c>0$ and $0<ac<\pi/4$, and choose
\[
0<\varepsilon<\min\{x/2,\ y/2,\ (x^4-y)/5\}.
\]
\begin{proposition}\label{prop:failure}
For every $n\geq1$,
\begin{equation}\label{eq:failure}
\Pr\{\text{\eqref{eq:est} is inadmissible}\}
\leq8\exp\!\{-\lfloor n/2\rfloor\varepsilon^2/2\}.
\end{equation}
\end{proposition}

The proof is an elementary application of concentration inequalities, and can be found in Appendix~\ref{app:failure}.
\end{samepage}

\section{Joint expansion and frequency choice}\label{sec:bias}

 In this section, we shall state results that explicitly compare the estimator of \cite{roll1984} with \eqref{eq:est}, and allow us subsequently to analyse combinations of the estimators.

\subsection{A Joint central limit theorem}\label{sec:jointclt}

 We first define the empirical version of Roll's estimator $\hat{c_R}$ and its leading order bias term $B_R$ by
\[
\begin{aligned}
\widehat\lambda_R&=-\frac1{n-1}\sum_{t=1}^{n-1}Y_tY_{t+1}, &
\widehat c_R&=\sqrt{\max(0,\widehat\lambda_R)},\\
L_R&=V^2+4Vc^2+5c^4, & B_R&=-\frac{L_R}{8c^3}.
\end{aligned}
\]
Write $\rho=V/c^2$ for the noise-to-spread variance ratio. Fix $c,V>0$ and frequencies $a_1,\ldots,a_J$ with $0<a_jc<\pi/4$. Let $b(a;c,V)$ and $\sigma_{\CF}^2(a;c,V)$ denote the leading CF bias and variance coefficients given in \eqref{eq:BK}--\eqref{eq:b} of Appendix~\ref{app:coefficients}. Abbreviate $b(a_j)=b(a_j;c,V)$, write $\boldsymbol1$ for the vector of $J+1$ ones, and set
\[
\begin{aligned}
\widehat{\boldsymbol c}
&=(\widehat c_R,\widehat c_{\CF}(a_1),\ldots,
\widehat c_{\CF}(a_J))^{\mathsf T},\\
\boldsymbol B&=(B_R,b(a_1),\ldots,b(a_J))^{\mathsf T}.
\end{aligned}
\]
For $0<z<\pi/4$, define
\begin{equation}\label{eq:Hdefinition}
H(z)=\frac{\tan z(2-\tan^2z)}{2z(1-\tan^2z)}.
\end{equation}
Let $\Sigma$ be the symmetric matrix with diagonal entries $\Sigma_{00}=L_R/(4c^2)$ and $\Sigma_{jj}=\sigma_{\CF}^2(a_j;c,V)$, $1\leq j\leq J$, and Roll--CF entries
\begin{equation}\label{eq:mixturecovariance}
\Sigma_{0j}=\Sigma_{j0}=\frac V2+c^2H(a_jc),\quad 1\leq j\leq J.
\end{equation}
The remaining entries are specified in Lemma~\ref{lem:mixturecovariance}. A joint central limit theorem for the estimators follows.
\begin{theorem}\label{thm:positive}
For these fixed parameters and frequencies, as $n\to\infty$,
\begin{equation}\label{eq:bias}
\begin{aligned}
\E\widehat{\boldsymbol c}-c\boldsymbol1
&=\boldsymbol B/n+O(n^{-2}),\\
\E\{(\widehat{\boldsymbol c}-c\boldsymbol1)
(\widehat{\boldsymbol c}-c\boldsymbol1)^{\mathsf T}\}
&=\Sigma/n+O(n^{-2}),
\end{aligned}
\end{equation}
and
\begin{equation}\label{eq:clt}
\sqrt n(\widehat{\boldsymbol c}-c\boldsymbol1)
\Rightarrow N(0,\Sigma).
\end{equation}
The remainders hold componentwise, uniformly on compact subsets of the admissible parameter and frequency domain.
\end{theorem}

The proof is in Appendices~\ref{app:positive} and~\ref{app:probability}, with the covariance calculations in Appendix~\ref{app:mixturecovariance}. The joint expansion provides the bias and covariance terms needed for aggregation with Roll.

\subsection{Modified versions and their comparison with Roll's estimator}\label{sec:rollfrequency}

Unlike Roll's coefficient $B_R<0$, the CF bias coefficient changes sign with frequency. Solving $b(a;c,V)=0$ analytically is difficult because \eqref{eq:b} combines exponential and trigonometric terms. Its endpoint behavior nevertheless guarantees a solution. For fixed $c>0$ and $V\geq0$, the small-frequency expansion in Appendix~\ref{app:smallfrequency} and the endpoint calculations \eqref{eq:smallbias} and~\eqref{eq:upperbias} in Appendix~\ref{app:root} give
\[
\begin{aligned}
\lim_{a\downarrow0}b(a;c,V)&<0,\\
\lim_{a\uparrow\pi/(4c)}b(a;c,V)&=+\infty.
\end{aligned}
\]
Thus $b$ lies below the horizontal axis near the left endpoint and above it near the right endpoint. Now continuity on the admissible interval guarantees an interior zero. We state this observation as a proposition.

\begin{proposition}\label{prop:root}
For fixed $c>0$ and $V\geq0$, there is a frequency $a_*$ satisfying
\begin{equation}\label{eq:root}
b(a_*;c,V)=0,\qquad 0<a_*<\frac{\pi}{4c}.
\end{equation}
At any such frequency, $\E\widehat c_{\CF}(a_*)-c=O(n^{-2})$.
\end{proposition}

\begin{figure}[H]
\centering
\includegraphics[width=\columnwidth]{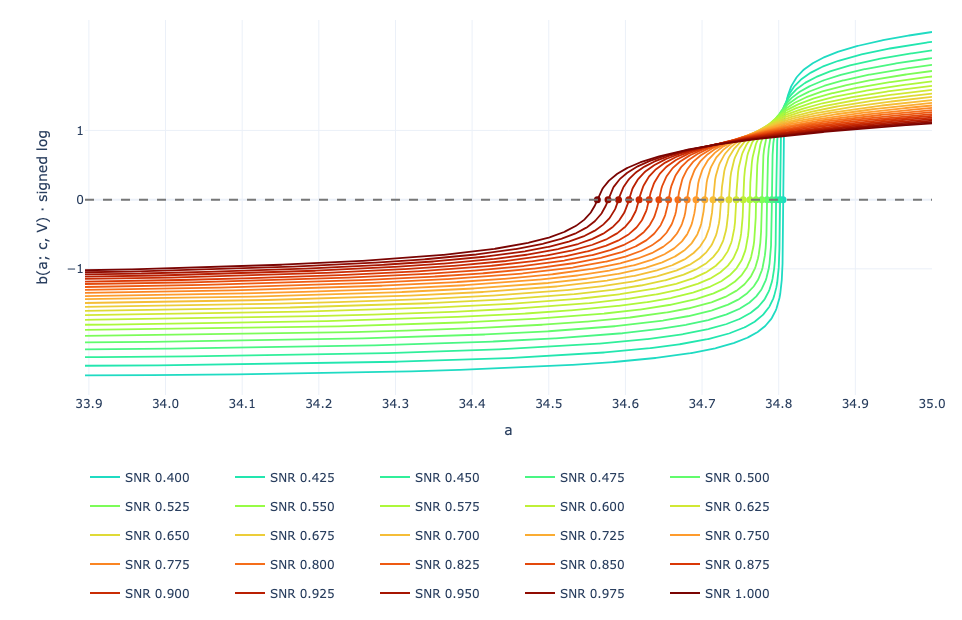}
\caption{Leading bias coefficient $b(a;c,V)$ for $c=0.02$ and $V=c^2/\mathrm{SNR}$, with $\mathrm{SNR}=0.400,0.425,\ldots,1.000$. The vertical axis uses $T(b)=\operatorname{sign}(b)\log_{10}(1+|b|/0.02)$. Dashed lines mark zero bias and the admissible endpoint $a=\pi/(4c)$. Zero crossings identify bias-neutral frequencies in Proposition~\ref{prop:root}.}
\label{fig:signed-log-bias-curve}
\end{figure}

The proof is in Appendix~\ref{app:root}. The opposite endpoint signs provide a bracket for numerical root finding, using the equivalent four-frequency equation \eqref{eq:computableroot} in Appendix~\ref{app:computableroot}. Bias cancellation leaves sampling error of order $n^{-1/2}$ when $V>0$, motivating the aggregation in Section~\ref{sec:aggregation}.
Figure~\ref{fig:signed-log-bias-curve} illustrates these sign changes and the dependence of the bias-neutral frequency on the signal-to-noise ratio.

Note that the bias cancellation above concerns fixed values of $c>0$. The following results compare CF and Roll at values near $0$. Fix $a,V>0$. Define $\widehat c_{\CF}^{+}(a)$ by replacing $\widehat d(a)$ in \eqref{eq:est} with $\max(0,\widehat d(a))$, using a fixed bounded value when the ratio is undefined. Let $s=Va^2$, $Z\sim N(0,\tau_a^2)$, and
\begin{equation}\label{eq:tau}
\tau_a^2=2\cosh(s)\{\cosh(s)-1\}^2\{\cosh(s)+2\}.
\end{equation}

\begin{theorem}\label{thm:boundary}
At $c=0$,
\begin{equation}\label{eq:boundary}
\sqrt n\,\widehat d(a)\Rightarrow Z,
n^{1/8}\widehat c_{\CF}^{+}(a)\Rightarrow a^{-1}\{\max(0,Z)\}^{1/4}.
\end{equation}
For $c_n=hn^{-1/8}$ and fixed $h\geq0$,
\begin{equation}\label{eq:local}
n^{1/8}\widehat c_{\CF}^{+}(a) \Rightarrow a^{-1}\{\max(0,a^4h^4+Z)\}^{1/4}.
\end{equation}
\end{theorem}
The proof is in Appendix~\ref{app:boundary}. For the Roll benchmark, retain $\widehat\lambda_R$, $\widehat c_R$, and $L_R$ from Section~\ref{sec:jointclt}; take $n\geq2$, $V>0$, and $Z_R\sim N(0,1)$.
\begin{theorem}\label{thm:roll}
For fixed $c\geq0$, $\E\widehat\lambda_R=c^2$ and
\begin{equation}\label{eq:rollclt}
\sqrt n(\widehat\lambda_R-c^2)\Rightarrow N(0,L_R).
\end{equation}
For fixed $c>0$,
\begin{equation}\label{eq:rollvar}
\Var(\widehat c_R)=\frac{L_R}{4c^2n}+o(n^{-1}).
\end{equation}
For $c_n=hn^{-1/4}$ and fixed $h\geq0$,
\begin{equation}\label{eq:rollboundary}
n^{1/4}\widehat c_R\Rightarrow\sqrt{\max(0,h^2+VZ_R)}.
\end{equation}
\end{theorem}
The proof is in Appendix~\ref{app:roll}. Thus, at $c=0$, the clipped CF estimator has the slower scale $n^{-1/8}$, compared with Roll's $n^{-1/4}$.
\FloatBarrier

\section{ECF--GMM and weighted mixtures with Roll}\label{sec:aggregation}

In this section, we discuss combinations of our estimator and Roll's estimator. There are two ways to combine characteristic-function information with Roll's lag-one moment. A convex mixture combines the spread estimates and can reduce both leading absolute bias and MSE. Augmented ECF--GMM (Empirical Characteristic Function -- Generalized Moment Method) combines the moments before estimation and has a leading variance comparison with Roll.

\subsection{The weighted CF--Roll estimator}
Since bias removal alone need not improve MSE, we combine the estimate at $a_*$ with Roll, using the same returns, to achieve both a bias and MSE reduction. Write the normalized covariance as
\[
G=c^{-2}\lim_{n\to\infty}n\Cov\{\widehat c_R,\widehat c_{\CF}(a_*)\}.
\]
Lemma~\ref{lem:mixturecovariance} gives this covariance explicitly through the function $H$ defined in \eqref{eq:Hdefinition}.
\begin{theorem}\label{thm:mixturegain}
Fix $a_*$ with $0<a_*c<\pi/4$ and $b(a_*;c,V)=0$. Let $R=L_R/(4c^4)$, and also let
\[
C=\sigma_{\CF}^2(a_*;c,V)/c^2,
D_*=C+R-2G>0.
\]
For $0\leq w\leq1$, define
$\widehat c_w=(1-w)\widehat c_R+w\widehat c_{\CF}(a_*)$. Then
\begin{equation}\label{eq:mixturebiasmse}
\begin{aligned}
\E\widehat c_w-c&=\frac{(1-w)B_R}{n}+O(n^{-2}),\\
\frac{n\E(\widehat c_w-c)^2}{c^2}
&=R-2w(R-G)+w^2D_*+O(n^{-1}).
\end{aligned}
\end{equation}
There exists $w\in(0,1]$ that strictly reduces both leading absolute
bias and leading MSE relative to Roll if and only if
\begin{equation}\label{eq:mixturegainregion}
R>G.
\end{equation}
In this case, one may choose any $w$ satisfying
\[
0<w\leq1,\qquad w<\frac{2(R-G)}{D_*}.
\]
The minimizing convex weight for the variance coefficient is
\begin{equation}\label{eq:optimalmixtureweight}
w_* =\min\{1,\max\{0,(R-G)/D_*\}\}.
\end{equation}
If $0<w_*<1$, the leading MSE and absolute bias ratios to Roll are,
respectively,
\[
1-\frac{(R-G)^2}{RD_*},\qquad 1-w_*.
\]
\end{theorem}
The proof of this result is in Appendix~\ref{app:mixturegain}. The condition $R>G$ warrants a few comments. It is seen to be strictly weaker than requiring $C<R$, and in experiments seems to hold moderately often. A large value of $C$ can still make the gain negligible: we refer to  Appendix~\ref{app:gaincondition} for the precise details.

\subsection{Augmented ECF--GMM with Roll}\label{sec:gmm}

We first define an empirical characteristic function-Generalized Moment method (ECF--GMM) for the marginal moments, and then augment its moment vector with Roll's moment. Specifically, on the parameter set $\zeta=(c,V)\in(0,\infty)^2$, we intend to define a loss function intended for minimization.

We require the following notation. Fix $J\geq2$ and positive frequencies $u_1,\ldots,u_J$. Let $\widehat f_j=\widehat\phi(u_j)$, and $
f_j(\zeta)= \phi(u_j)=e^{-Vu_j^2/2}\cos^2(cu_j)$ for $ j=1,\ldots,J$. Also define $\widehat f=(\widehat f_1,\ldots,\widehat f_J)^{\mathsf T}$, and $f(\zeta)=(f_1(\zeta),\ldots,f_J(\zeta))^{\mathsf T}$.

At the true parameter, we would like $\E\widehat f=f(\zeta)$. Thus, for a fixed positive definite matrix $W$, the weighted discrepancy
\begin{equation}\label{eq:gmmdefinition}
\begin{aligned}
\mathcal L_n(\zeta;W)
&=\{\widehat f-f(\zeta)\}^{\mathsf T}W\{\widehat f-f(\zeta)\}
\end{aligned}
\end{equation}
is a natural notion of loss. Let $\widehat\zeta_W=(\widehat c_G,\widehat V_G)$ be an interior local
minimizer of \eqref{eq:gmmdefinition}. We select a solution satisfying
$\widehat\zeta_W\overset{p}\longrightarrow\zeta$, as assumed in the results below.
Thus $\widehat c_G$ estimates the half spread and $\widehat V_G$ estimates
the innovation variance. The choice $W=I_J$ gives ordinary squared discrepancies. Other positive definite weights account for their different variances and covariances.

An augmented ECF--GMM estimator can now be specified as follows: we append the sample lag-one product moment $-\widehat\lambda_R$
to $\widehat f$, and its model value $-c^2$ to $f(\zeta)$.
The corresponding augmented derivative and long-run covariance
matrices can be used in the same weighted criterion. The latter includes the lagged covariances estimated in \eqref{eq:jointweight}. The next result shows that the empirical loss minimizers of \eqref{eq:gmmdefinition} are asymptotically Gaussian around the true value.

\begin{proposition}\label{prop:gmmweight}
Suppose $D=\partial f/\partial\zeta^{\mathsf T}$ has rank two and
\[
\Omega_{jk}=\frac{\phi(u_j-u_k)+\phi(u_j+u_k)}2
-\phi(u_j)\phi(u_k)
\]
is positive definite. For fixed positive definite $W$,
$\sqrt n(\widehat\zeta_W-\zeta)\Rightarrow N(0,\mathcal V(W))$, where
\begin{equation}\label{eq:gmmsandwich}
\mathcal V(W)=(D^{\mathsf T}WD)^{-1}D^{\mathsf T}W\Omega WD(D^{\mathsf T}WD)^{-1}.
\end{equation}
For every such $W$, $\mathcal V(W)-\mathcal V(\Omega^{-1})\succeq0$. Furthermore, for the augmented estimator, under the same rank and positivity conditions, optimal weighting gives a spread variance coefficient at most $L_R/(4c^2)$.
\end{proposition}

This is the variance comparison for augmented ECF--GMM; it does not
give a bias comparison. The scalar weight $w$ in
Theorem~\ref{thm:mixturegain} combines estimators, whereas $W$ here
weights moments. Appendix~\ref{app:gmmweight} contains the proof.

\section{Pilot selection}\label{sec:selection}
In this section, we describe a pilot-selection extension of the joint CF-Roll estimator discussed in Theorem~\ref{thm:mixturegain}. This is inspired by the construction in \cite{hansen1982}, and a similar construction for the augmented ECF-GMM estimator in the previous section can be found in Appendix~\ref{app:gmmweight}.

\subsection{A single sample pilot estimator}

Recall $Y_t$ from \eqref{yt} and consider the following quantities:
\[
\begin{aligned}
\widehat\gamma_0&=\frac1m\sum_{t=1}^mY_t^2,&
\widehat\gamma_1&=\frac1m\sum_{t=1}^mY_tY_{t+1},\\
\widetilde c_m^2&=-\widehat\gamma_1,&
\widetilde V_m&=\widehat\gamma_0+2\widehat\gamma_1.
\end{aligned}
\]
Here $\widetilde c_m^2$ is a raw estimate of $c^2$. Let $\widetilde c_m$ be its positive square root when that estimate is positive.

Suppose $a_*$ is a simple root of \eqref{eq:root}. On a fixed compact interval $I\subset(0,\pi/(4c))$ containing $a_*$, we solve
\begin{equation}\label{eq:pilotroot}
b(\widehat a_m;\widetilde c_m,\widetilde V_m)=0.
\end{equation}
We use a fixed point in $I$ when the local root is unavailable. Our main result is that the pilot estimator gives further bias reduction than the original ECF estimator.
\begin{theorem}\label{thm:pilot}
Under this construction, for $m,n\to\infty$,
\begin{equation}\label{eq:pilotclt}
\begin{aligned}
\widehat a_m-a_*&=O_p(m^{-1/2}),\\
\sqrt n\{\widehat c_{\CF}(\widehat a_m)-c\}
&\Rightarrow N(0,\sigma_{\CF}^2(a_*;c,V)),
\end{aligned}
\end{equation}
and
\begin{equation}\label{eq:pilotbias}
\begin{aligned}
\E\{\widehat c_{\CF}(\widehat a_m)\mid\mathcal P_m\}-c
&=O_p(n^{-1}m^{-1/2}+n^{-2}),
\end{aligned}
\end{equation}
where $\mathcal P_m$ is the $\sigma$-field generated by $\{Y_1,\ldots,Y_{m+1}\}$.
\end{theorem}
No restriction on the relative growth of $m$ and $n$ is needed.
The proof can be found in Appendix~\ref{app:pilot}.

In practice, we would reject non-positive pilot parameters and roots without adequate admissibility margins. Appendix~\ref{app:pilotselection} specifies a fallback and gives explicit concentration bounds that assert admissibility with high probability.

\subsection{Selecting an aggregate}\label{sec:aggregate}

Pilot uncertainty can make an apparent gain over Roll's model misleading. In this section, we discuss how to select values of $a$ and utilize Theorem~\ref{thm:positive} appropriately, such that a pilot construction succeeds with high probability. We also discuss when Roll's estimator may perform better.
 
 Write $\beta=(c^2,V)$ as the vector we wish to estimate. Let $\phi(a;\beta) = e^{-Va^2/2}\cos^2(ac)$. First, fix a value $m \geq 3$. We use the first $Y_{1},Y_2,\ldots,Y_{m+1}$ to form a pilot estimate $\widehat\beta_m=(\widetilde c_m^2,\widetilde V_m)$, as specified in the previous section. We skip the next value of $Y$ (i.e. $Y_{m+2}$), and reserve the rest of the samples for estimation: this ensures that the remaining variates are independent of those used in the pilot.
  For a tolerance parameter $0<\delta<1$, set $q_m=\sqrt{1224\log(12/\delta)/\lfloor m/3\rfloor}$. Consider the set
\[
\begin{aligned}
e_m&=\frac{q_m\widehat\gamma_0}{1-3q_m},\\
\mathcal C_m&=\{\beta':\|\beta'-\widehat\beta_m\|_\infty\leq e_m\}.
\end{aligned}
\]
The value of $e_m$ ensures $\mathbb P(\beta_m \in \mathcal{C}_m) \geq 1-\delta$, as we show while proving the main result of this section. We now check if $\mathcal{C}_m$ is not too large, by pre-choosing some compact $\mathcal K \subset (0,\infty)^2$ and seeing if $q_m < \frac 13$ and $\mathcal{C}_m \subset \mathcal K$ holds: if either does not, then the situation is classed as exceptional and we defer to Roll's estimator on the remaining samples.

In the typical situation, we obtain the pilot root $\hat{a_m}$ as in the previous section. The values of $a$ for which $\hat{c}_{\CF}(a)$ will be weighted together will all be positive multiples of $\hat{a}_m$. We choose those multiples $a$ of $\hat{a}_m$ for which, if $\beta'=(c'^2,V')\in\mathcal C_m$, then
\begin{equation}\label{eq:frequencysafety}
\begin{gathered}
0<a\sqrt{\theta'}<\pi/4, \text{ and} \\ 
\min\{x/2,y/2,(x^4-y)/5\}>\kappa>0,
\end{gathered}
\end{equation}
where $x = \phi(a; \beta'),y = \phi(a;2 \beta')$ in \eqref{eq:xy} are evaluated at $\beta'$, and $\kappa$ is some fixed tolerance threshold that controls inadmissibility. 

Next, using $\boldsymbol B,\Sigma$ from Theorem~\ref{thm:positive}, we choose $(\widehat\alpha,\widehat\gamma)$ to maximize $\gamma\in[0,1]$ over nonnegative weights summing to one, with at most $J_{\max}$ positive CF weights (here, $J_{\max}$ is chosen prior to this step), subject to
\begin{equation}\label{eq:robustselection}
\begin{aligned}
|\alpha^{\mathsf T}\boldsymbol B(\beta')|
&\leq(1-\gamma)|B_R(\beta')|,\\
\alpha^{\mathsf T}\Sigma(\beta')\alpha
&\leq(1-\gamma)\Sigma_{00}(\beta')
\end{aligned}
\end{equation}
for every $\beta'\in\mathcal C_m$. $\gamma$ indicates the leading absolute bias and MSE across plausible parameters, with $\alpha$ providing the weights over the aforementioned frequencies. Now, finally, we consider the weighted estimator
\begin{equation}\label{eq:adaptiveestimator}
\widehat c_{\mathrm A}=\widehat\alpha^{\mathsf T}\widehat{\boldsymbol c}
\end{equation}
on the independent estimation block, assigning zero to inadmissible CF components. We set $\widehat\gamma=0$ whenever Roll's estimator is returned.

\begin{algorithm}[H]
\caption{Selecting a certified CF–Roll aggregate}\label{alg:aggregate}
\algrenewcommand{\algorithmicrequire}{\textbf{Input:}}
\algrenewcommand{\algorithmicensure}{\textbf{Output:}}
\begin{algorithmic}[1]
\Require $Y_1,\ldots,Y_{m+n+2}$; $m\geq3$, $n\geq2$; $\delta,\mathcal K,\kappa,J_{\max}$; fixed grid $\mathcal A_0$, root multipliers $\mathcal R$, and frequency interval $\mathcal I$.
\Ensure Estimate $\widehat c_{\mathrm A}$ and certified gain $\widehat\gamma$.
\State $\mathcal P\gets(Y_1,\ldots,Y_{m+1})$
\State $\mathcal E\gets(Y_{m+3},\ldots,Y_{m+n+2})$; discard $Y_{m+2}$.
\State Compute $\widehat\beta_m$ and $q_m$ from $\mathcal P$.
\If{$q_m\geq1/3$}
    \State \Return $(\operatorname{Roll}(\mathcal E),0)$
\EndIf
\State Form $e_m$ and $\mathcal C_m$ as above.
\If{$\mathcal C_m\not\subset\mathcal K$}
    \State \Return $(\operatorname{Roll}(\mathcal E),0)$
\EndIf
\State $\mathcal A\gets\mathcal A_0$
\If{a pilot root $\widehat a_m$ exists}
    \State $\mathcal A\gets\mathcal A\cup\{r\widehat a_m:r\in\mathcal R\}$
\EndIf
\State Retain in $\mathcal A$ only frequencies in $\mathcal I$ satisfying \eqref{eq:frequencysafety} for every $\beta'\in\mathcal C_m$.
\If{$\mathcal A=\varnothing$}
    \State \Return $(\operatorname{Roll}(\mathcal E),0)$
\EndIf
\ForAll{$S\subseteq\mathcal A$ with $|S|\leq J_{\max}$}
    \State Solve for maximal $\gamma_S\in[0,1]$ under \eqref{eq:robustselection} throughout $\mathcal C_m$, with simplex weights $\alpha_S$ on Roll and $S$.
\EndFor
\State Select $(\widehat\alpha,\widehat\gamma)$ with the largest gain, using fixed measurable tie-breaking; assign excluded frequencies zero weight.
\If{$\widehat\gamma=0$}
    \State \Return $(\operatorname{Roll}(\mathcal E),0)$
\EndIf
\State Freeze frequencies $a_1,\ldots,a_J$ and weights $\widehat\alpha$.
\State $z_1,\ldots,z_J\gets0$
\ForAll{$j\in\{1,\ldots,J\}$ with $\widehat\alpha_j>0$}
    \If{$\widehat c_{\CF}(a_j)$ is admissible on $\mathcal E$}
        \State $z_j\gets\widehat c_{\CF}(a_j)$ computed on $\mathcal E$.
    \EndIf
\EndFor
\State \Return $\bigl(\widehat\alpha_0\operatorname{Roll}(\mathcal E)+\sum_{j=1}^J\widehat\alpha_jz_j,\widehat\gamma\bigr)$
\Statex \textit{Keep the frozen weights when $z_j=0$; do not renormalize.}
\end{algorithmic}
\end{algorithm}

We are ready to state our main result of this section: our estimator comes with a \emph{certified} gain over Roll's model in both leading bias and MSE coefficients. The gain is certified by the parameter $\hat{\gamma}$.

\begin{theorem}\label{thm:adaptivecertificate}
If $\beta\in\mathcal K$, then with pilot probability at least $1-\delta$,
\begin{equation}\label{eq:adaptiveexpansions}
\begin{aligned}
|\E(\widehat c_{\mathrm A}\mid\mathcal P_m)-c|
&\leq\frac{(1-\widehat\gamma)|B_R|}{n}+O(n^{-2}),\\
\E\{(\widehat c_{\mathrm A}-c)^2\mid\mathcal P_m\}
&\leq\frac{(1-\widehat\gamma)\Sigma_{00}}{n}+O(n^{-2}).
\end{aligned}
\end{equation}
The remainders are uniform. A positive gain in both leading coefficients is falsely certified with probability at most $\delta$.
\end{theorem}

We provide an extremely brief proof sketch. First, observe that the procedure defers to Roll on a set that we have claimed to be of low probability: this is established in Theorem~\ref{thm:pilotconcentration}. Then, we apply a Taylor-expansion based argument, as explained in Appendix~\ref{app:probability}, to explain the origin of the terms $\mathbf{B}$ and $\Sigma$. The rest of the argument is rather straightforward. 

The proof and finite-sample conditions are in Appendices~\ref{app:adaptivecertificate}--\ref{app:finitemargin}.

\section{Experiments}\label{sec:simulations}

In this section, we discuss the experiments validating our theoretical results. First, we discuss an experiment performed on synthetic data. Subsequently, we compare our estimator with Roll's and Hasbrouck's estimator using India equities. Finally, we discuss the running time comparison of the algorithms.

\subsection{Synthetic experiments}

We compare Hasbrouck's estimator, Roll's estimator, and the augmented ECF--GMM estimator from Theorem~\ref{thm:pilotconcentration} on $1,000$ simulated $252$-day price paths for each displayed signal-to-noise ratio. The paths follow Roll's model with true half-spread $c=0.02$, independent trade signs $Q_t=\pm1$ with probability $1/2$ each, and Gaussian innovation variance $V$ varied with the signal-to-noise ratio. Each estimator uses the simulated prices without knowing the true $c$; the known value is used only to evaluate estimation error.
Hasbrouck's estimator uses 1,000 burn-in iterations and 1,500 retained draws in these experiments.

\begin{figure}[ht]
\centering
\includegraphics[width=\columnwidth]{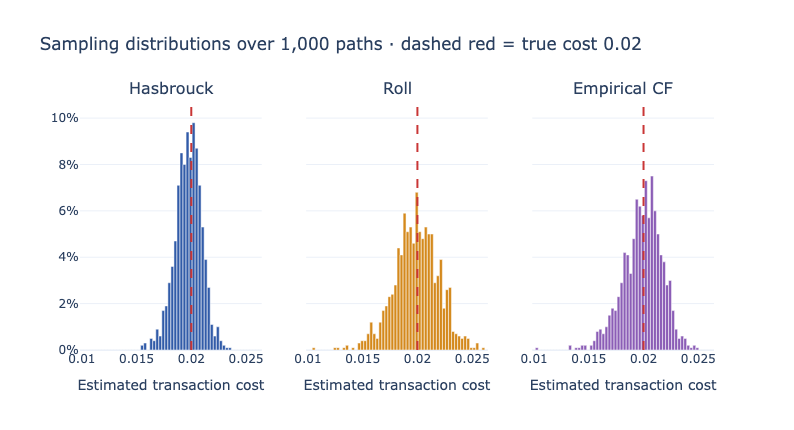}
\caption{Sampling distributions over 1,000 simulated 252-day paths under Roll's model. The dashed red line marks the true half-spread $c=0.02$, unknown to the estimators. Bar heights give percentages of paths; ``Empirical CF'' denotes augmented ECF--GMM.}
\label{fig:synthetic-distributions}
\end{figure}

Figure~\ref{fig:synthetic-distributions} shows the empirical sampling
distributions of the three estimates over 1,000 simulated price paths.
Hasbrouck's distribution is slightly shifted to the left of the true
value. The augmented ECF--GMM estimator is more dispersed, but its center is
closer to the true transaction cost. This shows that our estimator demonstrates significantly less variance than Roll's estimator at a similar computational cost, but loses out significantly to Hasbrouck's estimator.

\begin{figure}[h]
\centering
\includegraphics[width=\columnwidth]{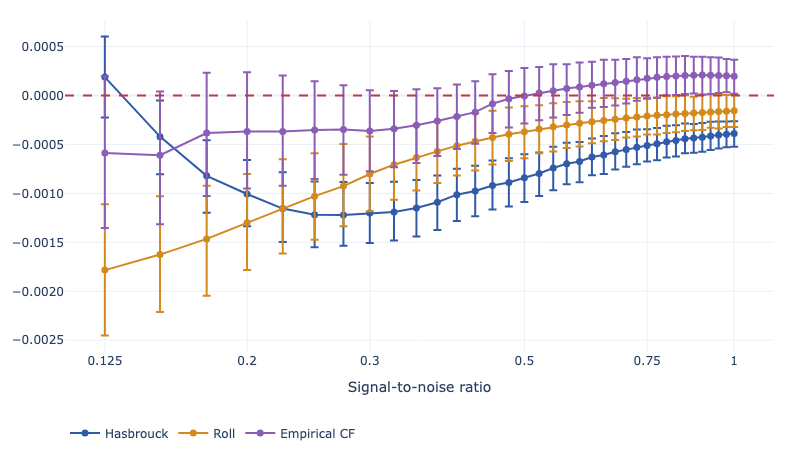}
\par\medskip
\includegraphics[width=\columnwidth]{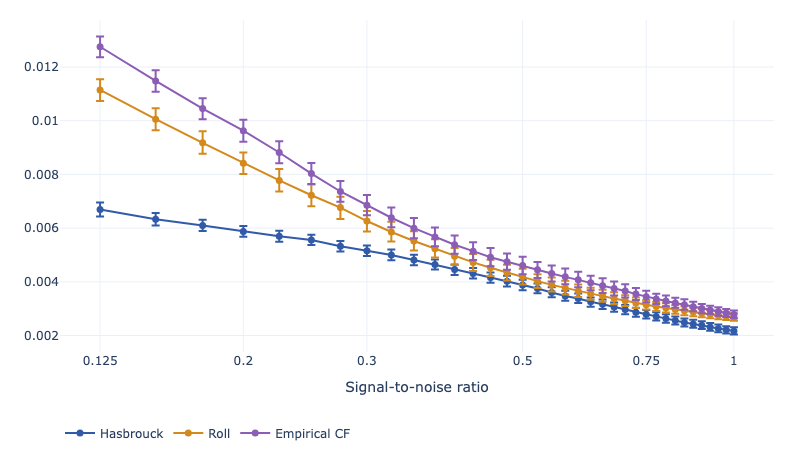}
\caption{Bias (top) and RMSE (bottom) across 1,000 simulated 252-day paths at each signal-to-noise ratio, with true $c=0.02$. Bias is the mean estimate minus $0.02$; RMSE is the square root of the mean squared estimation error. Error bars are 95\% confidence intervals. ``Empirical CF'' denotes augmented ECF--GMM.}
\label{fig:synthetic-bias-rmse}
\end{figure}

Figure~\ref{fig:synthetic-bias-rmse} compares bias and RMSE as the signal-to-noise ratio varies over the values shown, with the same model and true transaction cost. The CF estimates are closer to zero bias over much of the displayed range, yet their root MSE remains above the other two methods. These panels illustrate why reducing bias alone need not improve overall estimation accuracy.

\subsection{Indian equities}

\begin{figure}[!htbp]
\centering
\includegraphics[width=\columnwidth]{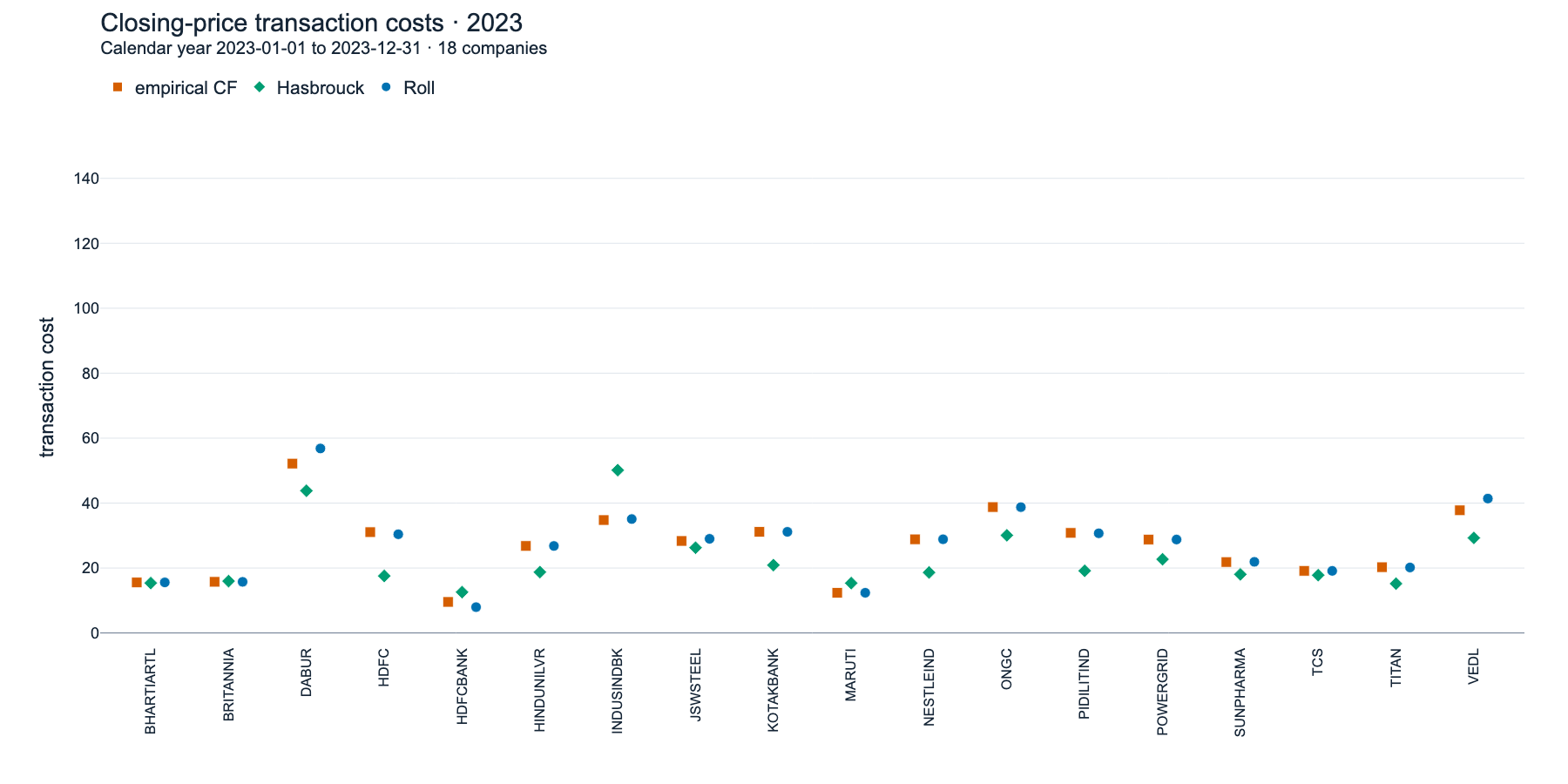}
\par\medskip
\includegraphics[width=\columnwidth]{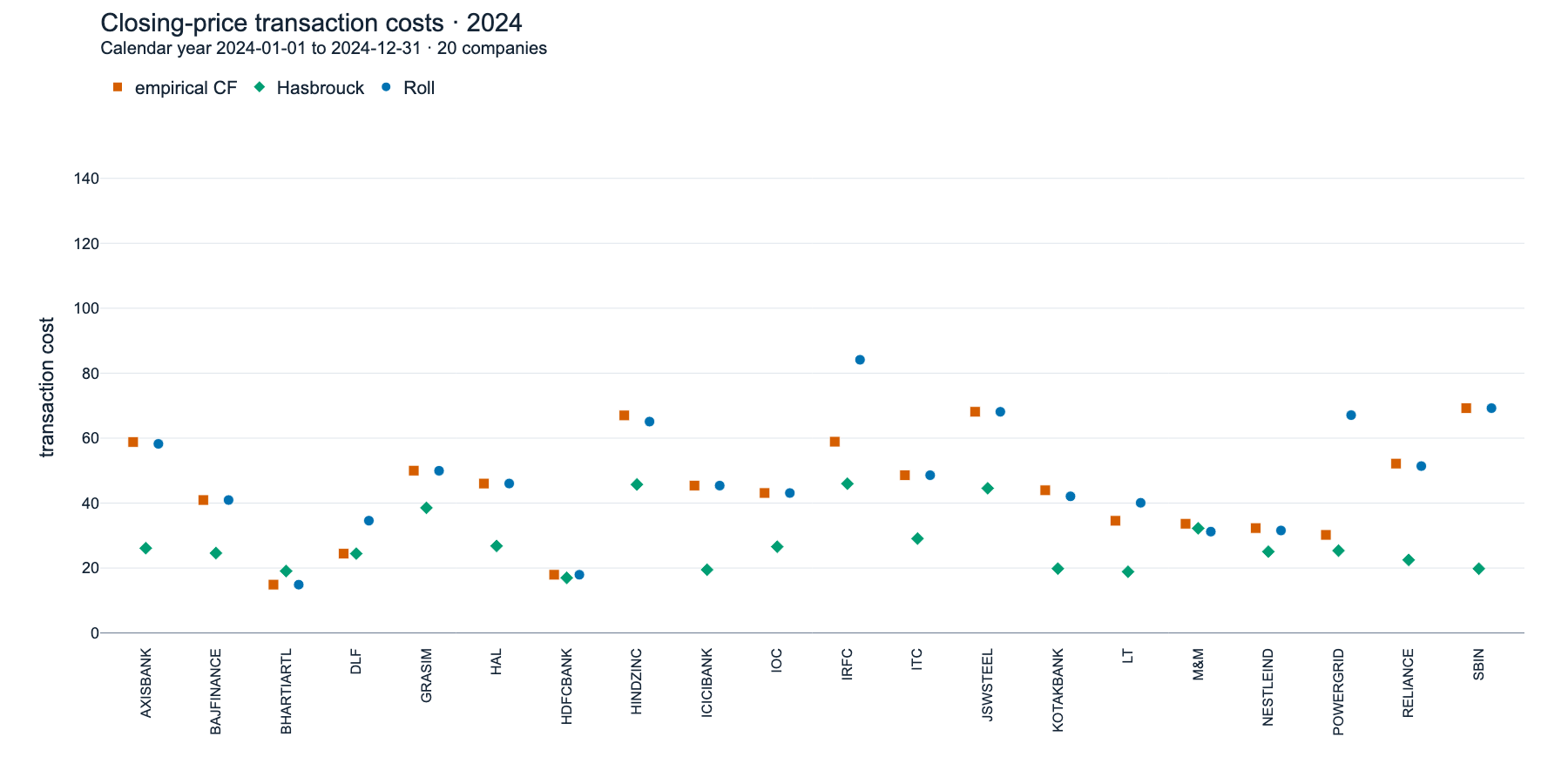}
\caption{Transaction-cost (log basis points) estimates for calendar year 2023 (top, 18 companies) and 2024 (bottom, 20 companies), using CMIE Prowess MCap20 data and retaining tickers with 252 trading-day closing prices. ``Empirical CF'' denotes augmented ECF--GMM.}
\label{fig:indian-equities}
\end{figure}

We use CMIE Prowess closing prices for the top-20 Indian equity universe by market capitalization in 2023 and 2024, retaining only tickers with 252 trading-day closes in each year's CSV. Similarly, Figure~\ref{fig:indian-equities} compares the methods for 18 retained companies in 2023 and 20 in 2024. We find that while the CF and Roll estimators often agree closely, the Hasbrouck estimator gives lower estimates for many stocks. 

\FloatBarrier

\subsection{Benchmarking computation time}

We now compare our estimator's running time with \cite{roll1984} and \cite{hasbrouck2009}. We report mean elapsed time per path over 1,000 simulated 252-day paths under the setup in Section~\ref{sec:simulations}, with true half-spread $c=0.02$ and variance varied with SNR. All runs used CPUs only on a system with $2\times$ Intel Xeon Gold 6530 processors. The Empirical CF timing includes the full execution of Algorithm~\ref{alg:aggregate}, using a separate, larger pilot sample.

\begin{table}[H]
\centering
\caption{Mean runtime per path}\label{tab:runtime}
\begin{tabular}{@{}lrr@{}}
\toprule
Method & Time (s) \\
\midrule
Roll & 0.000276 \\
Augemented ECF--GMM & 0.005769 \\
Hasbrouck & 39.525235 \\
\bottomrule
\end{tabular}
\end{table}

Augmented ECF--GMM takes $5.769$ ms per path, compared with $39.525235$ s for Hasbrouck, which is $6{,}851$ times faster. Roll is fastest at $0.276$ ms, about $20.9$ times faster than Augmented ECF--GMM. Thus, our estimator runs significantly faster than Hasbrouck's model. We provide a brief explanation using computational complexity.

For the leading per-path complexity, let $n$ be the number of returns in an estimation path. Roll needs one pass, so its cost is $O(n)$. For augmented ECF--GMM with six frequencies, empirical moments are accumulated once and reused during optimization; with $I$ optimizer iterations, we obtain the running time $O(n+I)$. For a varying number $J$ of frequencies, constructing the CF moments alone costs $O(nJ)$. Finally, a Hasbrouck Gibbs sweep updates the latent trade signs and parameter sufficient statistics in $O(n)$ work, giving $O(Tn)$ for $T$ sweeps per chain. The experimental settings give $T=1000+1500=2500$, including burn-in and retained draws. These repeated sweeps introduce a far higher computational multiplier, since our estimator works with just $J=4$. While this analysis does not include pilot selection, a similarly straightforward analysis still shows that the leading coefficient of $n$ in the running time is far exceeded by that for Hasbrouck's estimator.

\section{Conclusion}\label{sec:discussion}

We derive the leading \(n^{-1}\) bias and joint covariance of a two-frequency ECF spread estimator under the Gaussian Roll model, and show that an interior frequency choice eliminates the leading ECF bias for $c>0$. The joint expansion further identifies conditions under which a convex CF–Roll aggregate improves the leading absolute-bias and MSE coefficients relative to Roll. All our statements are conditionally asymptotic and do not imply uniform (over $c$) or substantial finite-sample dominance, as our experiments show.

While our estimator was designed explicitly for bias reduction, it is unclear if it is the best among some reasonable class of estimators at this task. It does not dominate Roll at small frequencies, where its variance and absolute bias can be much larger when the noise dominates the spread, as Theorems~\ref{thm:boundary} and \ref{thm:roll} show.

Furthermore, while we present comparisons with Hasbrouck's model \citep{hasbrouck2009}, we have not considered a similar bias reduction-based estimator which improves on Hasbrouck's work. Along with attempts to address these in future work, we also plan to integrate the estimator into various financial trading strategies to test its real-life performance.

\subsection*{AI use statement}

We used generative AI tools for language editing, manuscript organization, notation-consistency checks, high-level feedback on exposition, and experimental reporting. The tools did not generate the mathematical proofs, experimental code, or data. All technical claims, derivations, implementations, numerical outputs, and final interpretations were designed, independently checked and approved by the authors. The authors take full responsibility for the final content.

\appendix
\thispagestyle{empty}

\onecolumn

\section{Literature Review}\label{app:litreview}\label{sec:literature}
Spread estimators separate trading costs from efficient price movements. Much of this literature investigates moment based methods. Most daily estimators use prices alone. Intraday methods may also require quotes and trade directions.

\citet{roll1984} identifies the spread from the negative first order covariance of price changes. The estimator is simple and inexpensive. It relies on assumptions about efficient prices and trade signs. A positive sample covariance gives no real valued spread estimate. Setting such estimates to zero can introduce bias.

\citet{hasbrouck2009} uses Bayesian inference and Gibbs sampling to estimate costs from daily prices. The method allows nonnegative estimates when Roll's sample covariance is positive. It requires prior choices and repeated simulation. Priors can strongly influence estimates in short samples.

\citet{corwinschultz2012} use daily highs and lows to separate spread and volatility effects. They compare price ranges over one and two days. The method needs reliable extrema and adjustments for overnight movements. Negative estimates also require a reporting rule.

\citet{abdiranaldo2017} combine closing prices with daily highs and lows. Their method accommodates overnight price changes and performs well for less liquid stocks. It still relies on assumptions about spread and volatility stability. Estimated squared spreads can be negative.

\citet{jahanparvarzikes2023} evaluate these measures in equity and foreign exchange markets. They find substantial volatility effects when spreads are small relative to price variation. This makes bias correction important when interpreting changes in estimated trading costs.

\citet{ardiaguidottikroencke2024} combine opening, high, low, and closing prices to reduce estimation variance. Their estimator accounts for discrete price observations. It uses OHLC records and can retain upward bias when daily spreads are very small.

Intraday benchmarks compare transaction prices with prevailing quote midpoints \citep{hasbrouck2009}. They require reliable matching of trades and quotes. \citet{huangstoll1997} further separate order processing, inventory, and adverse selection costs. Their decomposition needs trade direction data and assumptions about how recorded trades relate to underlying orders.

We build a characteristic function based estimator from price changes. We then investigate bias correction techniques to improve its accuracy. The analysis gives an explicit bias formula. This supports direct subtraction and frequency selection using an independent pilot sample. Both target the leading bias at positive spreads. We also study variance and behavior near zero spread, since reducing bias alone need not improve overall accuracy.

\section{Technical lemmas and admissibility}\label{app:technical}\label{app:support}

\subsection{Dependence and Taylor bounds}\label{app:probability}
A sequence is $q$ dependent if collections separated by more than $q$ dates
are independent. The returns in \eqref{yt} are $1$ dependent. Vectors
built from $(Y_t,Y_{t+1})$ are $2$ dependent. Independence of innovations and
signs includes $Q_0$.

Let $X_t$ be a centered, stationary, $q$ dependent vector with fixed $q$ and
bounded fourth moments, and write $\overline X_n=n^{-1}\sum_tX_t$. Then
\begin{equation}\label{eq:app-moments}
\E\|\overline X_n\|^2=O(n^{-1}),\qquad
\E\|\overline X_n\|^4=O(n^{-2}),\qquad
\E(\overline X_{n,i}\overline X_{n,j}\overline X_{n,k})=O(n^{-2}).
\end{equation}
To check these orders, expand the products into sums over dates. A date
separated from all the others contributes zero by centering and independence.
There are $O(n)$ remaining triples and $O(n^2)$ remaining quadruples.
H\"older's inequality bounds each term. The variance uses only the $2q+1$
possible covariance lags.

The corresponding CLT is
\begin{equation}\label{eq:app-clt}
\sqrt n\,\overline X_n\Rightarrow N(0,\Sigma),\qquad
\Sigma=\sum_{h=-q}^{q}\Cov(X_0,X_h).
\end{equation}
One proof uses blocks of length $\ell_n=\lfloor n^{1/3}\rfloor$, separated
by $q$ omitted dates. Retained blocks are independent. The omitted sum has
variance $O(n/\ell_n+\ell_n)=o(n)$. Block fourth moments are $O(\ell_n^2)$,
so the Lindeberg bound is $O(\ell_n/n)=o(1)$. This also proves the triangular
array version when fourth moments are uniformly bounded and the long run
covariance converges. Zero variance directions follow by Chebyshev's inequality.

For $Z_t\in[-1,1]$, split the dates into $q+1$ independent residue classes.
Hoeffding's inequality and a union bound give
\begin{equation}\label{eq:app-concentration}
\Pr\left\{\left|\frac1n\sum_{t=1}^n(Z_t-\E Z_t)\right|>\eta\right\}
\leq2(q+1)e^{-\lfloor n/(q+1)\rfloor\eta^2/2}.
\end{equation}
For $n<q+1$ the bound is trivial.

Let $F$ be four times continuously differentiable near the population mean
$\mu$ of bounded observations, and bounded elsewhere. A third degree Taylor
expansion, \eqref{eq:app-moments}, and \eqref{eq:app-concentration} give
\begin{equation}\label{eq:app-taylor}
\E F(\mu+\overline X_n)
=F(\mu)+\frac12\sum_{i,j}F_{ij}(\mu)
\E(\overline X_{n,i}\overline X_{n,j})+O(n^{-2}).
\end{equation}
The expected cubic term and fourth order remainder are both $O(n^{-2})$.
Outside the smooth neighborhood, boundedness and concentration give an
exponentially small contribution. The same argument applies to second moments
and to smooth bounded extensions of intermediate ratios. Constants are uniform
on compact sets inside the smooth domain, but need not stay bounded near
zero spread or an admissibility boundary.

\subsection{Proof of Lemma~\ref{lem:identification}}\label{app:identification}
\begin{proof}
Independence and symmetry give
\[
\E e^{iuY_t}
=\E e^{iucQ_t}\E e^{-iucQ_{t-1}}\E e^{iuU_t}
=e^{-Vu^2/2}\cos^2(cu)=\phi(u).
\]
Since $0<ac<\pi/4$, we have $\cos(2ac)>0$ and
\[
1-\frac{\sqrt{\phi(2a)}}{\phi(a)^2}
=1-\frac{\cos(2ac)}{\cos^4(ac)}
=\frac{\sin^4(ac)}{\cos^4(ac)}=\tan^4(ac).
\]
The nonnegative fourth root is $\tan(ac)$, and the principal inverse tangent
returns $ac$. This proves \eqref{eq:identity}, with $x>0$ and $0<y<x^4$.
\end{proof}

\subsection{Proof of Lemma~\ref{lem:covariance}}\label{app:covariance}

\begin{lemma}\label{lem:covariance}
For every $n\geq1$,
\begin{equation}\label{eq:cov}
\Cov\begin{pmatrix}\widehat\phi(a)\\\widehat\phi(2a)\end{pmatrix}
=\frac1n\begin{pmatrix}S_{11}&S_{12}\\S_{12}&S_{22}\end{pmatrix},
\end{equation}
where
\begin{equation}\label{eq:entries}
\begin{aligned}
S_{11}&=\{1+\phi(2a)\}/2-x^2,\\
S_{12}&=\{\phi(a)+\phi(3a)\}/2-xy,\\
S_{22}&=\{1+\phi(4a)\}/2-y^2.
\end{aligned}
\end{equation}
\end{lemma}

\begin{proof}
Adjacent returns share only $Q_t$ and are
conditionally independent given that sign. For $q\in\{-1,1\}$,
\[
\begin{aligned}
\E\{\cos(uY_t)\mid Q_t=q\}
&=e^{-Vu^2/2}\cos(cu)\cos(cuq)=\phi(u),\\
\E\{\cos(vY_{t+1})\mid Q_t=q\}&=\phi(v).
\end{aligned}
\]
Thus $\Cov\{\cos(uY_t),\cos(vY_{t+1})\}=0$ for all $u,v$. Returns at larger
lags are independent. At the same date, the cosine product identity gives
\[
\Cov\{\cos(uY_t),\cos(vY_t)\}
=\tfrac12\{\phi(u-v)+\phi(u+v)\}-\phi(u)\phi(v).
\]
Only the $n$ same date terms remain in the covariance of the sample averages.
Substituting $(u,v)=(a,a),(a,2a),(2a,2a)$ proves
\eqref{eq:cov} to \eqref{eq:entries} for every $n$.
\end{proof}

\subsection{Proof of Proposition~\ref{prop:failure}}\label{app:failure}
\begin{proof}
Apply \eqref{eq:app-concentration} with $q=1$ at both frequencies.
Except on an event of probability at most
$8e^{-\lfloor n/2\rfloor\varepsilon^2/2}$, we have
\[
|\widehat\phi(a)-x|\leq\varepsilon,\qquad
|\widehat\phi(2a)-y|\leq\varepsilon.
\]
On this event, $\widehat\phi(a)>x/2>0$ and
$\widehat\phi(2a)>y/2>0$. Since $v\mapsto v^4$ has derivative at most
four on $[0,1]$,
\[
\widehat\phi(a)^4-\widehat\phi(2a)
\geq x^4-y-4|\widehat\phi(a)-x|-|\widehat\phi(2a)-y|
\geq x^4-y-5\varepsilon>0.
\]
These inequalities ensure admissibility. For $n=1$, the bound exceeds one.
\end{proof}

\section{Proofs of the main expansion and bias-neutral frequency}\label{app:mainexpansion}\label{app:proofs}

\subsection{Explicit CF coefficients}\label{app:coefficients}
For the CF estimate, put
$r=\sqrt y/x^2$, $d=1-r$, and
\begin{equation}\label{eq:BK}
\begin{aligned}
B&=r\left(-\frac{3S_{11}}{x^2}+\frac{S_{12}}{xy}+\frac{S_{22}}{8y^2}\right),\\
K&=r^2\left(\frac{4S_{11}}{x^2}-\frac{2S_{12}}{xy}+\frac{S_{22}}{4y^2}\right).
\end{aligned}
\end{equation}
The variance and bias coefficients are
\begin{equation}\label{eq:cfvariance}
\sigma_{\CF}^2(a;c,V)=\frac{K}{16a^2d^{3/2}(1+\sqrt d)^2},
\end{equation}
\begin{equation}\label{eq:b}
b(a;c,V)=\frac{B}{4a d^{3/4}(1+\sqrt d)}
\quad-\frac{(3+5\sqrt d)K}{32a d^{7/4}(1+\sqrt d)^2}.
\end{equation}

\subsection{Proof of Theorem~\ref{thm:positive}}\label{app:positive}
\begin{proof}
Let $S=(S_{ij})_{i,j=1}^2$ and
$\Delta_n=(\widehat\phi(a)-x,\widehat\phi(2a)-y)^{\mathsf T}$.
Lemma~\ref{lem:covariance} and the finite dependence CLT in
Section~\ref{app:probability} give
\[
\E\Delta_n=0,\qquad \E(\Delta_n\Delta_n^{\mathsf T})=S/n,
\qquad \sqrt n\Delta_n\Rightarrow N(0,S).
\]
Use the ratio and inverse maps
\[
g(v,w)=1-\sqrt w/v^2,\qquad h_a(z)=a^{-1}\arctan(z^{1/4}).
\]
At $(x,y)$, direct differentiation gives
\[
\nabla g=\begin{pmatrix}2r/x\\-r/(2y)\end{pmatrix},\qquad
\nabla^2g=\begin{pmatrix}-6r/x^2&r/(xy)\\r/(xy)&r/(4y^2)\end{pmatrix},
\]
\[
h_a'(d)=\frac1{4a d^{3/4}(1+\sqrt d)},\qquad
h_a''(d)=-\frac{3+5\sqrt d}{16a d^{7/4}(1+\sqrt d)^2}.
\]
Hence
\[
\tfrac12\sum_{i,j}g_{ij}S_{ij}=B,\qquad
(\nabla g)^{\mathsf T}S\nabla g=K.
\]
The delta method gives variance $h_a'(d)^2K$, as in \eqref{eq:cfvariance}.
It is positive for $V>0$: $Y_t$ has an everywhere positive density, and no
nonzero linear combination of its two cosine transforms is constant.

The population point lies strictly inside the smooth domain. The estimator
is bounded by $\pi/(4a)$ on admissible samples. Its fixed bounded extension
and the concentration bound therefore allow \eqref{eq:app-taylor} to be used.
The chain rule yields
\[
\begin{aligned}
\E\widehat c_{\CF}(a)-c
&=\frac1{2n}\sum_{i,j}
\{h_a'(d)g_{ij}+h_a''(d)g_i g_j\}S_{ij}+O(n^{-2})\\
&=\frac{h_a'(d)B+\tfrac12h_a''(d)K}{n}+O(n^{-2}).
\end{aligned}
\]
Substituting the derivatives gives \eqref{eq:b} and the CF bias expansion
in \eqref{eq:bias}.
The same argument holds at $V=0$, where $S$ may be singular.

For the joint statement, stack the cosine moments at all $a_j$ with
$\widehat\lambda_R$. The resulting centered moment vector has finite
dependence and moments of every order under the Gaussian model.
The finite dependence CLT in Appendix~\ref{app:probability} therefore
applies to every linear combination. The CF coordinates have the smooth
maps used above, and the Roll coordinate has derivative $1/(2c)$ at
$\lambda_R=c^2$. The tail and moment bounds in Appendix~\ref{app:roll}
control its truncation outside this neighborhood.

Appendix~\ref{app:rawbias} gives the Roll bias coefficient $B_R$.
Appendix~\ref{app:mixturecovariance} gives the cross covariances.
Expand each coordinate to third order. Mixed cubic expectations and fourth
moments are $O(n^{-2})$ by the finite dependence counting argument.
The linear terms give $\Sigma/n$, while all remaining terms in the joint
second moment are $O(n^{-2})$. The squared mean is also $O(n^{-2})$.
This proves both joint expansions and the multivariate normal limit.
The bounds are uniform on compact parameter and frequency sets strictly
inside the admissible domain.
\end{proof}

\subsection{Proof of Proposition~\ref{prop:root}}\label{app:root}
\begin{proof}
Put $\rho=V/c^2$. Equations~\eqref{eq:app:smallcoefficients} and~\eqref{eq:app:Wvalue} give
\begin{equation}\label{eq:smallbias}
\lim_{a\downarrow0}\frac{b(a;c,V)}c
=-\frac{9\rho^4}{64}-\frac{9\rho^3}{8}-\frac{21\rho^2}{16}+\frac{3\rho}{4}-\frac34<0,
\end{equation}
The right hand side is
\[
 -\frac9{64}\rho^4-\frac98\rho^3
 -\frac3{16}\{7(\rho-2/7)^2+24/7\}<0
 \qquad(\rho\geq0).
\]
At the other endpoint, let $a\uparrow a_0=\pi/(4c)$. Then $y\downarrow0$, $d\to1$, and
\[
 x\to x_0=\tfrac12e^{-Va_0^2/2},\qquad
 S_{22}\to s_0=\tfrac12(1+e^{-8Va_0^2}).
\]
Both limits are positive. The other covariance entries stay bounded. Thus
\begin{equation}\label{eq:upperbias}
\begin{aligned}
 B&=\frac{S_{22}}{8x^2y^{3/2}}+O(y^{-1/2}),\\
 K&=\frac{S_{22}}{4x^4y}+O(1),\\
 b(a;c,V)&\sim\frac{s_0}{64a_0x_0^2}\,y^{-3/2}\to+\infty.
\end{aligned}
\end{equation}
The denominators of \eqref{eq:b} are positive in the interior, so $b$ is continuous there. The intermediate value theorem supplies a root, and \eqref{eq:bias} gives bias $O(n^{-2})$ at that root. This also holds at $V=0$: the bounded observations, finite dependence, and interior Taylor expansion used in Appendix~\ref{app:positive} remain valid.
\end{proof}

\subsection{The computable frequency equation}\label{app:computableroot}

The bias-neutral frequency equation uses model values at only four frequencies. Set $p_k=\phi(ka)$ and
\[
\begin{aligned}
A&=\frac{1+p_2}{2p_1^2}-1,&
D&=\frac{p_1+p_3}{2p_1p_2}-1,\\
E&=\frac{1+p_4}{2p_2^2}-1.
\end{aligned}
\]
Thus $A=S_{11}/x^2$, $D=S_{12}/(xy)$, and $E=S_{22}/y^2$.

With $t=\tan^2(ac)\in(0,1)$, the equation $b(a;c,V)=0$ is equivalent to
\begin{equation}\label{eq:computableroot}
(9t^2-2t-3)E=16(t^2+2t+3)A-8(3-t)(1+t)D.
\end{equation}

Put $t=\tan^2(ac)\in(0,1)$. Since $d=t^2$, $r=1-t^2$,
$B=r(-3A+D+E/8)$, and $K=r^2(4A-2D+E/4)$, the equation $b=0$ is equivalent to
\[
 8t^2(1+t)(-3A+D+E/8)
 \qquad-(3+5t)(1-t^2)(4A-2D+E/4)=0.
\]
Dividing by $1+t$ and collecting coefficients gives \eqref{eq:computableroot}. Every removed factor is positive.

\section{Covariances and proofs for the two mixtures}\label{app:aggregation}\label{app:weightedproofs}

\subsection{Proof of Lemma~\ref{lem:mixturecovariance}}\label{app:mixturecovariance}

\begin{lemma}\label{lem:mixturecovariance}
The Roll variance and Roll--CF covariances are given by
\eqref{eq:Hdefinition}--\eqref{eq:mixturecovariance}.
For $i,j\geq1$, $\Sigma_{ij}=q_i^{\mathsf T}K_{ij}q_j$, where
\[
\begin{aligned}
(K_{ij})_{k\ell}
&=\frac{\phi(ka_i-\ell a_j)+\phi(ka_i+\ell a_j)}2\\
&\quad-\phi(ka_i)\phi(\ell a_j),\qquad k,\ell\in\{1,2\},\\
q_j&=\frac{1}{4a_jd_j^{3/4}(1+\sqrt{d_j})}
\begin{pmatrix}2r_j/\phi(a_j)\\-r_j/(2\phi(2a_j))\end{pmatrix},\\
r_j&=\frac{\sqrt{\phi(2a_j)}}{\phi(a_j)^2},\qquad d_j=1-r_j.
\end{aligned}
\]
\end{lemma}

\begin{proof}
Write $W_t=Y_tY_{t+1}$. Conditional expectations given the shared signs,
together with $\E\{U\sin(uU)\}=Vu e^{-Vu^2/2}$, yield
\begin{equation}\label{eq:cosrollcross}
\sum_{k\in\mathbb Z}\Cov\{\cos(uY_0),W_k\}
=e^{-Vu^2/2}\{2c^2\sin^2(cu)+cVu\sin(2cu)\}=:\Xi(u).
\end{equation}
Only $k=-1,0$ contribute. To check one term, condition on $(Q_{-1},Q_0)$:
\[
\begin{aligned}
\E\{\cos(uY_0)W_0\}
&=-c\,\E\{Q_0Y_0\cos(uY_0)\}\\
&=-c^2e^{-Vu^2/2}\cos(2cu)
+\tfrac12cVu e^{-Vu^2/2}\sin(2cu),\\
\Cov\{\cos(uY_0),W_0\}
&=e^{-Vu^2/2}\{c^2\sin^2(cu)+\tfrac12cVu\sin(2cu)\}.
\end{aligned}
\]
The term for $W_{-1}$ is equal by the corresponding conditioning. The outer
overlaps have zero covariance because the conditional cosine mean is constant.
Applying the CF gradient and the Roll derivative $-1/(2c)$ gives
\[
\Sigma_{0j}=-\frac{q_{j,1}\Xi(a_j)+q_{j,2}\Xi(2a_j)}{2c}.
\]
Substitute $z=a_jc$, $\tan(2z)=2\tan z/(1-\tan^2z)$, and
$r_j=1-\tan^4z$ to obtain $V/2+c^2H(z)$.
The remaining entries follow from Lemma~\ref{lem:covariance} and the
multivariate delta method. Finite endpoints change only terms of order
$n^{-2}$ in covariances of the moment averages.
\end{proof}

\subsection{Proof of Theorem~\ref{thm:mixturegain}}\label{app:mixturegain}
\begin{proof}
Linearity of expectation and $b(a_*;c,V)=0$ give the bias formula.
Lemma~\ref{lem:mixturecovariance} gives
\[
c^{-2}\lim_n n\Var(\widehat c_w)
=(1-w)^2R+2w(1-w)G+w^2C
=R-2w(R-G)+w^2D_*.
\]
The squared bias contributes $O(n^{-2})$ to MSE. For $w>0$, the variance
difference from Roll is negative exactly when $wD_*<2(R-G)$.
Existence of such a $w\leq1$ is equivalent to $R>G$, which is
\eqref{eq:mixturegainregion}. Every positive convex weight reduces
$|(1-w)B_R|$. Minimizing the quadratic gives
\eqref{eq:optimalmixtureweight}. Substitution gives the MSE ratio.
Uniform Taylor bounds give the stated remainders.
\end{proof}

\subsection{Interpreting the gain condition}\label{app:gaincondition}
rite $z=a_*c$ and $\rho=V/c^2$. Since
\[
H(z)=\frac{\tan z}{z}
\left(1+\frac{\tan^2z}{2(1-\tan^2z)}\right)>1
\qquad(0<z<\pi/4),
\]
the definitions of $R$ and $G$ give the exact equivalence
\begin{equation}\label{eq:gainthreshold}
\begin{aligned}
R-G&=\frac{(\rho+1)^2-4\{H(z)-1\}}4,\\
R>G&\quad\Longleftrightarrow\quad
\rho>2\sqrt{H(z)-1}-1.
\end{aligned}
\end{equation}
This is a threshold conditional on $z$; the bias-neutral root itself depends on $\rho$. The covariance inequality $G\leq\sqrt{RC}$ shows that $C<R$ implies $R>G$, while the converse need not hold. Thus the mixture can improve on Roll even when CF alone has larger leading variance.

Solving \eqref{eq:computableroot} numerically gives
\[
\begin{array}{c|rrr}
\rho & z & R-G & C\\\hline
1 & 0.67855232 & -0.29191267 & 20.91545256\\
2 & 0.69126736 & \phantom{-}0.75321182 & 181.44371515
\end{array}
\]
These examples establish neither a universal cutoff in $\rho$ nor uniqueness of the root. At the displayed root for $\rho=2$, \eqref{eq:optimalmixtureweight} gives $w_*\approx0.004215$, and the fractional reduction in leading MSE is approximately $0.000747$ ($0.0747\%$). Hence satisfying $R>G$ need not deliver a substantial gain.

\subsection{Proof of Proposition~\ref{prop:gmmweight}}\label{app:gmmweight}

The estimating equations are
\[
D(\widehat\zeta_W)^{\mathsf T}W
\{\widehat f-f(\widehat\zeta_W)\}=0,
\]
where $D=\partial f/\partial\zeta^{\mathsf T}$ has entries
\[
\begin{aligned}
D_{j1}(c,V)&=-u_j e^{-Vu_j^2/2}\sin(2cu_j),\\
D_{j2}(c,V)&=-\tfrac12u_j^2e^{-Vu_j^2/2}\cos^2(cu_j).
\end{aligned}
\]
\paragraph{Independent pilot weights.}
For an independent pilot of $m$ returns, define
$\widehat\phi_m(u)=m^{-1}\sum_{t=1}^m\cos(uY_t)$. Its covariance estimate is
\begin{equation}\label{eq:empiricalweight}
\widehat\Omega_{jk}=\frac{\widehat\phi_m(u_j-u_k)+\widehat\phi_m(u_j+u_k)}2\quad-\widehat\phi_m(u_j)\widehat\phi_m(u_k).
\end{equation}
Clip its eigenvalues to $[d_-,d_+]\subset(0,\infty)$ and invert.
If every eigenvalue of $\Omega$ lies in $(d_-,d_+)$, this pilot weight
attains $\mathcal V(\Omega^{-1})$ as $m,n\to\infty$.

For augmented ECF--GMM, use $m+1$ pilot returns and set
\[
\begin{aligned}
Z_t&=\bigl(\cos(u_1Y_t),\ldots,\cos(u_JY_t),Y_tY_{t+1}\bigr)^{\mathsf T},\\
\overline Z&=m^{-1}\sum_{t=1}^mZ_t.
\end{aligned}
\]
Replace \eqref{eq:empiricalweight} by
\begin{equation}\label{eq:jointweight}
\begin{aligned}
\widehat\Omega&=\widehat\Gamma_0+
\sum_{h=1}^2(\widehat\Gamma_h+\widehat\Gamma_h^{\mathsf T}),\\
\widehat\Gamma_h&=\frac1m\sum_{t=1}^{m-h}
(Z_t-\overline Z)(Z_{t+h}-\overline Z)^{\mathsf T},
\end{aligned}
\end{equation}
then apply the same clipping and inversion.

\begin{proof}
Lemma~\ref{lem:covariance} and the finite dependence CLT imply
$\sqrt n(\widehat f-f)\Rightarrow N(0,\Omega)$. The normal equations give
\[
\sqrt n(\widehat\zeta_W-\zeta)
=A_W\sqrt n(\widehat f-f)+o_p(1),\qquad
A_W=(D^{\mathsf T}WD)^{-1}D^{\mathsf T}W.
\]
Let $A_*=(D^{\mathsf T}\Omega^{-1}D)^{-1}D^{\mathsf T}\Omega^{-1}$.
Since $A_WD=A_*D=I$,
\[
\mathcal V(W)-\mathcal V(\Omega^{-1})
=(A_W-A_*)\Omega(A_W-A_*)^{\mathsf T}\succeq0.
\]
The cosine product identity makes \eqref{eq:empiricalweight} the empirical
covariance of the cosine vector. The finite dependence law of large numbers
gives $\widehat\Omega\to\Omega$. Continuity of eigenvalue clipping and
inversion proves consistency of the weight when clipping is asymptotically
inactive. Independence of the pilot and Slutsky's theorem complete the claim.

For the augmented vector $Z_t$, the model mean is $(f(\zeta)^{\mathsf T},-c^2)^{\mathsf T}$. Dependence extends at most two lags, so
\eqref{eq:jointweight} consistently estimates its long run covariance.
The spread influence coefficient using only the last moment is
$A_R=(0,\ldots,0,-1/(2c))$. For the augmented derivative $D$,
$A_RD=(1,0)$. If $A_{*,c}$ is the spread row of the efficient influence matrix,
\[
A_R\Omega A_R^{\mathsf T}-A_{*,c}\Omega A_{*,c}^{\mathsf T}
=(A_R-A_{*,c})\Omega(A_R-A_{*,c})^{\mathsf T}\geq0.
\]
This is the variance comparison with Roll. No bias comparison follows from it.
\end{proof}

\section{Pilot selection and concentration}\label{app:pilotselection}

\subsection{Moment estimates for the pilot}\label{app:pilot-moments}
Under the Gaussian model,
\[
\E Y_t^2=2c^2+V,\qquad \E(Y_tY_{t+1})=-c^2.
\]
The two sample averages have errors $O_p(m^{-1/2})$ by finite dependence and
finite fourth moments. Hence
\[
\widetilde c_m^2=-\widehat\gamma_1,\qquad
\widetilde V_m=\widehat\gamma_0+2\widehat\gamma_1
\]
are root $m$ consistent. Taking the square root preserves this rate at $c>0$,
and both estimates are positive with probability tending to one when $c,V>0$.

\subsection{Proof of Theorem~\ref{thm:pilot}}\label{app:pilot}
\begin{proof}
Let $I$ be the stated compact neighborhood of $a_*$.
Since $\partial_a b(a_*;c,V)\ne0$, the implicit function theorem gives
\[
\widehat a_m-a_*
=-\frac{\partial_c b(a_*;c,V)(\widetilde c_m-c)
       +\partial_V b(a_*;c,V)(\widetilde V_m-V)}
       {\partial_a b(a_*;c,V)}
+o_p(m^{-1/2})=O_p(m^{-1/2}).
\]
Use a fixed fallback in $I$ when the local root is unavailable. This event
has probability tending to zero.

A pilot ending at $Y_{m+1}$ uses signs through $Q_{m+1}$. An estimation block
starting at $Y_{m+3}$ uses signs from $Q_{m+2}$ onward. Their innovations
are also disjoint, so the two blocks are independent.
Let $\mathcal P_m$ denote the pilot sigma field.

On $I$, the positive margins $x(a)$, $y(a)$, and $x(a)^4-y(a)$ are bounded
away from zero. The bias expansion is therefore uniform, giving
\[
\E\{\widehat c_{\CF}(\widehat a_m)\mid\mathcal P_m\}-c
=\frac{b(\widehat a_m;c,V)}n+O(n^{-2})
=O_p(n^{-1}m^{-1/2}+n^{-2}).
\]
Here we used $b(a_*;c,V)=0$ and boundedness of $\partial_a b$ on $I$.

For the limiting distribution, use the maps from Section~\ref{app:positive}
and define
\[
\ell_a(Y)=h_a'(d(a))\nabla g(x(a),y(a))^{\mathsf T}
\begin{pmatrix}\cos(aY)-x(a)\\\cos(2aY)-y(a)\end{pmatrix}.
\]
Because $h_a(g(x(a),y(a)))=c$ for every $a\in I$, Taylor expansion gives
\[
\sqrt n\{\widehat c_{\CF}(\widehat a_m)-c\}
=\frac1{\sqrt n}\sum_{t=1}^n\ell_{\widehat a_m}(Y_t)+o_p(1).
\]
The functions $\ell_a$ are uniformly bounded and continuous in $L^2$.
Lemma~\ref{lem:covariance} also gives zero cross date covariance at any two
frequencies. Thus, conditionally on the pilot,
\[
\Var\left\{\frac1{\sqrt n}\sum_{t=1}^n
[\ell_{\widehat a_m}(Y_t)-\ell_{a_*}(Y_t)]\,\middle|\,\mathcal P_m\right\}
=\E\{\ell_a(Y)-\ell_{a_*}(Y)\}^2\big|_{a=\widehat a_m}=o_p(1).
\]
Boundedness and Chebyshev's inequality make this difference $o_p(1)$.
The fixed frequency CLT at $a_*$ proves \eqref{eq:pilotclt}, with no
restriction on the relative growth of $m$ and $n$.
\end{proof}

\subsection{Proof of Theorem~\ref{thm:pilotconcentration}}\label{app:pilotconcentration}

Use $m+1$ pilot returns and leave one unused return before the estimation
block. For $\beta=(\theta,V)$, $\theta=c^2$, and $T=V+2\theta$, define
\begin{equation}\label{eq:pilotbetadefinition}
\begin{aligned}
\widetilde c_m^2&=-\frac1m\sum_{t=1}^mY_tY_{t+1},&
\widehat\gamma_0&=\frac1m\sum_{t=1}^mY_t^2,\\
\widetilde V_m&=\widehat\gamma_0-2\widetilde c_m^2,&
\widehat\beta_m&=(\widetilde c_m^2,\widetilde V_m).
\end{aligned}
\end{equation}

\begin{theorem}\label{thm:pilotconcentration}
Under \eqref{yt}, let $m\geq3$ and $C_0=1/1224$. For every $t>0$,
\begin{equation}\label{eq:pilotconcentration}
\begin{gathered}
\Pr\{\|\widehat\beta_m-\beta\|_\infty>t\}\leq p_m(t,T),\\
p_m(t,T)=\min\{1,12e^{-C_0\lfloor m/3\rfloor\min(t^2/T^2,t/T)}\}.
\end{gathered}
\end{equation}
For $0<\delta<1$, define
\begin{equation}\label{eq:pilotconfidence}
\begin{aligned}
q_m&=\sqrt{\frac{\log(12/\delta)}{C_0\lfloor m/3\rfloor}},\\
e_m&=\frac{q_m\widehat\gamma_0}{1-3q_m},\\
\mathcal C_m&=\{\beta':\|\beta'-\widehat\beta_m\|_\infty\leq e_m\}.
\end{aligned}
\end{equation}
Then $q_m<1/3$ implies $\Pr(\beta\in\mathcal C_m)\geq1-\delta$.
\end{theorem}

\begin{proof}
The Gaussian innovation and the signs give
\[
\E e^{sY_t}=e^{Vs^2/2}\cosh^2(cs)\leq e^{Ts^2/2},\qquad
\E|Y_t|^{2k}\leq2(2T)^k k!\quad(k\geq1).
\]
For $X_t=Y_t^2$ or $X_t=Y_tY_{t+1}$, the arithmetic geometric mean
inequality and centering imply
\[
\E|X_t-\E X_t|^k\leq2(4T)^k k!
=\frac{k!}{2}(64T^2)(4T)^{k-2}\quad(k\geq2).
\]
For $\ell$ independent centered terms satisfying this bound, the exponential
moment argument for Bernstein's inequality \citep{vershynin2018} gives
\[
\Pr\{|\overline X-\E X|>t\}
\leq2\exp\left\{-\frac{\ell t^2}{128T^2+8Tt}\right\}
\leq2\exp\left\{-\frac\ell{136}
\min\left(\frac{t^2}{T^2},\frac tT\right)\right\}.
\]
Split each pilot sum into the three residue classes modulo three. Terms in
each class are independent. If each class average is within $t$ of its
mean, so is their weighted average. Hence each pilot average has tail at
most six times the last exponential with $\ell=\lfloor m/3\rfloor$.
The implication
\[
|\widetilde c_m^2-\theta|\leq t/3,\quad
|\widehat\gamma_0-T|\leq t/3
\quad\Longrightarrow\quad
\|\widehat\beta_m-\beta\|_\infty\leq t
\]
and a union bound prove \eqref{eq:pilotconcentration}, with $C_0=1/(9\cdot136)$.

For $q_m<1/3$, the event
$\|\widehat\beta_m-\beta\|_\infty\leq q_mT$ has probability at least
$1-\delta$. On this event,
\[
|\widehat\gamma_0-T|\leq3q_mT,\qquad
T\leq\frac{\widehat\gamma_0}{1-3q_m},\qquad
\|\widehat\beta_m-\beta\|_\infty\leq e_m.
\]
This proves coverage.
\end{proof}

\subsection{Proof of Theorem~\ref{thm:adaptivecertificate}}\label{app:adaptivecertificate}

In this section, we restate a more complete version of Theorem~\ref{thm:adaptivecertificate} and prove it.  Recall Algorithm~\ref{alg:aggregate}.

\begin{theorem}\label{thm:adaptivecertificatefull}
Assume the true parameter lies in the prescribed compact set and use
\eqref{eq:frequencysafety} and \eqref{eq:robustselection}.
With pilot probability at least $1-\delta$,
\begin{equation}\label{eq:adaptivecertificate}
\begin{aligned}
|\widehat\alpha^{\mathsf T}\boldsymbol B(\beta)|
&\leq(1-\widehat\gamma)|B_R(\beta)|,\\
\widehat\alpha^{\mathsf T}\Sigma(\beta)\widehat\alpha
&\leq(1-\widehat\gamma)\Sigma_{00}(\beta).
\end{aligned}
\end{equation}
On the same event, conditional on the pilot,
\begin{equation}
\begin{aligned}
|\E(\widehat c_{\mathrm A}\mid\mathcal P_m)-c|
&\leq\frac{(1-\widehat\gamma)|B_R|}{n}+O(n^{-2}),\\
\E\{(\widehat c_{\mathrm A}-c)^2\mid\mathcal P_m\}
&\leq\frac{(1-\widehat\gamma)\Sigma_{00}}{n}+O(n^{-2}).
\end{aligned}
\end{equation}
The remainder constants are uniform. The probability of falsely
certifying a positive gain in both leading coefficients is at most $\delta$.

Let $\gamma^*$ be the largest common fractional gain at $\beta$ over the
same retained frequencies and size constraint. Suppose both ratios
\[
\frac{|\alpha^{\mathsf T}\boldsymbol B(\beta')|}{|B_R(\beta')|},
\qquad
\frac{\alpha^{\mathsf T}\Sigma(\beta')\alpha}{\Sigma_{00}(\beta')}
\]
have Lipschitz constant at most $L$ in the maximum norm on
$\mathcal C_m\cup\{\beta\}$, uniformly over allowed $\alpha$.
On $\{\beta\in\mathcal C_m\}$, whenever selection is attempted,
\begin{equation}\label{eq:adaptiveoracle}
0\leq\gamma^*-\widehat\gamma\leq2Le_m.
\end{equation}
For $0<\varepsilon<\kappa$,
\begin{equation}\label{eq:adaptivefailure}
\Pr\!\left\{
\exists j\geq1:\widehat\alpha_j>0,
\widehat c_{\CF}(a_j)\text{ is inadmissible}
\right\}\leq\delta+8J_{\max}e^{-\lfloor n/2\rfloor\varepsilon^2/2}.
\end{equation}
\end{theorem}

\begin{proof}
If $q_m\geq1/3$, the procedure returns Roll and the coefficient and failure
claims are immediate. Otherwise, let $\mathcal E_m=\{\beta\in\mathcal C_m\}$.
Theorem~\ref{thm:pilotconcentration} gives
$\Pr(\mathcal E_m)\geq1-\delta$. On $\mathcal E_m$, feasibility of
\eqref{eq:robustselection} at $\beta'=\beta$ proves
\eqref{eq:adaptivecertificate}. A fallback to Roll satisfies the same
inequalities with $\widehat\gamma=0$.

Conditional on the pilot, the frequencies and weights are fixed and the
estimation block is independent. On the compact admissible domain, the
Taylor argument in Appendix~\ref{app:probability}, applied jointly to all
cosine moments and the Roll product, gives uniformly
\[
\begin{aligned}
\E(\widehat{\boldsymbol c}\mid\mathcal P_m)
&=c\boldsymbol1+\boldsymbol B/n+O(n^{-2}),\\
\E\{(\widehat{\boldsymbol c}-c\boldsymbol1)
(\widehat{\boldsymbol c}-c\boldsymbol1)^{\mathsf T}\mid\mathcal P_m\}
&=\Sigma/n+O(n^{-2}).
\end{aligned}
\]
For the unbounded Roll coordinate, Gaussian moment bounds and the Bernstein
tail bound justify the same expansion. Finite dependence bounds all cubic
expectations by $O(n^{-2})$ and fourth moments by $O(n^{-2})$.
Multiplication by the simplex weights proves \eqref{eq:adaptiveexpansions}.

Let $g(\alpha,\beta')$ be one minus the maximum of the two relative criteria
in the theorem, and let $\mathcal D_m$ be the allowed simplex with the size
constraint. Then
\[
\gamma^*=\max_{\alpha\in\mathcal D_m}g(\alpha,\beta),\qquad
\widehat\gamma=\max_{\alpha\in\mathcal D_m}
\inf_{\beta'\in\mathcal C_m}g(\alpha,\beta').
\]
Both maxima are nonnegative because Roll is feasible. On $\mathcal E_m$,
every $\beta'\in\mathcal C_m$ is within $2e_m$ of $\beta$. Thus
\[
g(\alpha,\beta)-2Le_m
\leq\inf_{\beta'\in\mathcal C_m}g(\alpha,\beta')
\leq g(\alpha,\beta).
\]
Maximizing proves \eqref{eq:adaptiveoracle}. Existence follows by compactness
and continuity, since the size constrained simplex is a finite union of
closed faces.

For every selected frequency on $\mathcal E_m$, condition
\eqref{eq:frequencysafety} implies the premise of
Proposition~\ref{prop:failure} for $\varepsilon<\kappa$.
Condition on the pilot and apply a union bound over at most $J_{\max}$
frequencies. Adding $\Pr(\mathcal E_m^c)$ proves \eqref{eq:adaptivefailure}.
No bound over unselected frequencies is required.
\end{proof}

\subsection{Proof of Proposition~\ref{prop:finitemargin}}\label{app:finitemargin}

\begin{proposition}\label{prop:finitemargin}
Suppose the bias and MSE remainders for both Roll and the selected
estimator are bounded in absolute value by $K_b/n^2$ and $K_M/n^2$,
respectively, with uniform constants. On $\{\beta\in\mathcal C_m\}$,
\[
\begin{aligned}
n\widehat\gamma\inf_{\beta'\in\mathcal C_m}|B_R(\beta')|&>2K_b,\\
n\widehat\gamma\inf_{\beta'\in\mathcal C_m}\Sigma_{00}(\beta')&>2K_M
\end{aligned}
\]
imply
\[
\begin{aligned}
|\E(\widehat c_{\mathrm A}\mid\mathcal P_m)-c|
&<|\E\widehat c_R-c|,\\
\E\{(\widehat c_{\mathrm A}-c)^2\mid\mathcal P_m\}
&<\E(\widehat c_R-c)^2.
\end{aligned}
\]
\end{proposition}

\begin{proof}
On the coverage event, the bias difference is bounded below by
\[
|\E\widehat c_R-c|
-|\E(\widehat c_{\mathrm A}\mid\mathcal P_m)-c|
\geq\frac{\widehat\gamma|B_R(\beta)|}{n}-\frac{2K_b}{n^2}.
\]
The MSE difference is at least
$\widehat\gamma\Sigma_{00}(\beta)/n-2K_M/n^2$.
Each lower bound is positive under the corresponding displayed condition.
\end{proof}

\section{Boundary behavior and comparisons}\label{app:comparisons}

\subsection{Proof of Theorem~\ref{thm:boundary}}\label{app:boundary}

\begin{proof}
Use $g$ from Section~\ref{app:positive}. At $c=0$, put
$s=Va^2$, $x_0=e^{-s/2}$, $y_0=e^{-2s}$, and let $S_0$ denote
\eqref{eq:entries} at these parameters. Then
\[
\nabla g(x_0,y_0)=\left(2/x_0,-1/(2y_0)\right)^{\mathsf T},\qquad
\begin{aligned}
(S_0)_{11}/x_0^2&=\cosh(s)-1,\\
(S_0)_{12}/(x_0y_0)&=\cosh(2s)-1,\\
(S_0)_{22}/y_0^2&=\cosh(4s)-1.
\end{aligned}
\]
The variance is
\[
\begin{aligned}
\tau_a^2
&=\nabla g(x_0,y_0)^{\mathsf T}S_0\nabla g(x_0,y_0)\\
&=4\{\cosh(s)-1\}-2\{\cosh(2s)-1\}
  +\tfrac14\{\cosh(4s)-1\}\\
&=2\cosh^4(s)-6\cosh^2(s)+4\cosh(s)\\
&=2\cosh(s)\{\cosh(s)-1\}^2\{\cosh(s)+2\}>0.
\end{aligned}
\]

For $c_n=hn^{-1/8}$, define
$X_{nt}=(\cos(aY_{nt}),\cos(2aY_{nt}))^{\mathsf T}$,
$(x_n,y_n)^{\mathsf T}=\E X_{nt}$, and
$\Delta_n=\overline X_n-(x_n,y_n)^{\mathsf T}$. These bounded vectors are one dependent, have
zero lag one covariance by Lemma~\ref{lem:covariance}, and have
same date covariance $S_n\to S_0$. Section~\ref{app:probability}
therefore gives $\sqrt n\Delta_n\Rightarrow N(0,S_0)$.
Since $(x_n,y_n)\to(x_0,y_0)$ with positive coordinates,
$g$ has bounded second derivatives on a common neighborhood. Taylor
expansion yields
\[
\begin{aligned}
\sqrt n\{\widehat d(a)-g(x_n,y_n)\}
&=\nabla g(x_n,y_n)^{\mathsf T}\sqrt n\Delta_n
  +O_p(n^{-1/2})\Rightarrow Z,\\
\sqrt n\,g(x_n,y_n)
&=\sqrt n\tan^4(ac_n)=a^4h^4+O(n^{-1/4}),
\qquad Z\sim N(0,\tau_a^2).
\end{aligned}
\]
The concentration bound in Section~\ref{app:probability} makes the
complement of this neighborhood exponentially unlikely, so the bounded
extension does not affect these limits.

Since $\sqrt n\,\widehat d(a)\Rightarrow a^4h^4+Z$ and
$\arctan z=z+O(z^3)$,
\[
\begin{aligned}
n^{1/8}\widehat c_{\CF}^{+}(a)
&=\frac{n^{1/8}}a\arctan\!\left(n^{-1/8}\{\max(0,\sqrt n\,\widehat d(a))\}^{1/4}\right)\\
&=a^{-1}\{\max(0,\sqrt n\,\widehat d(a))\}^{1/4}+O_p(n^{-1/4})\\
&\Rightarrow a^{-1}\{\max(0,a^4h^4+Z)\}^{1/4}.
\end{aligned}
\]
Continuity of $v\mapsto\{\max(0,v)\}^{1/4}$ proves \eqref{eq:local}.
The case $h=0$ gives \eqref{eq:boundary}. With probability tending to one,
both cosine averages are positive, so failure of the unclipped
estimator is equivalent to $\widehat d(a)<0$. At $c=0$, its limiting
probability is $\Pr(Z<0)=1/2$.
\end{proof}

\subsection{Proof of Theorem~\ref{thm:roll}}\label{app:roll}

\begin{proof}
Put $D_t=Q_t-Q_{t-1}$ and $W_t=Y_tY_{t+1}$.
The identities $\E D_t=0$, $\E D_t^2=2$, and
$\E(D_tD_{t+1})=-1$ give $\E W_t=-c^2$, hence
$\E\widehat\lambda_R=c^2$.

The products $W_t$ are two dependent. Conditional on $Q_t$,
$D_t$ and $D_{t+1}$ are independent with second moments $2$, so
$\E(D_t^2D_{t+1}^2)=4$. Integrating out the innovations gives
\[
\E W_t^2
=\E\{(V+c^2D_t^2)(V+c^2D_{t+1}^2)\}
=V^2+4Vc^2+4c^4.
\]
For lag one, $\E(D_tD_{t+2})=0$. Conditioning on $(Q_t,Q_{t+1})$ gives
\[
\begin{aligned}
\E(D_tD_{t+1}^2D_{t+2})
 &=\E\{-Q_tQ_{t+1}(Q_{t+1}-Q_t)^2\}=\E(2-2Q_tQ_{t+1})=2,\\
\E(W_tW_{t+1})
 &=c^2V\E(D_tD_{t+2})+c^4\E(D_tD_{t+1}^2D_{t+2})=2c^4.
\end{aligned}
\]
At lag two, $W_t$ and $W_{t+2}$ are conditionally independent given
their only shared sign $Q_{t+1}$, and both conditional means equal
$-c^2$. Thus, writing $\gamma_W(j)=\Cov(W_t,W_{t+j})$,
\[
\gamma_W(0)=V^2+4Vc^2+3c^4,\qquad
\gamma_W(1)=c^4,\qquad \gamma_W(j)=0\quad(j\geq2).
\]
It follows that
\[
L_R=\gamma_W(0)+2\gamma_W(1)=V^2+4Vc^2+5c^4,
\qquad
\Var(\widehat\lambda_R)=\frac{L_R}{n-1}-\frac{2c^4}{(n-1)^2}.
\]

All fourth moments of $W_t$ are finite. The central limit theorem
in Section~\ref{app:probability}, with $(n-1)/n\to1$, proves
\[
\sqrt n(\widehat\lambda_R-c^2)\Rightarrow N(0,L_R).
\]
For $c>0$, $\widehat\lambda_R\to c^2$ in probability. The delta
method with derivative $1/(2c)$ therefore gives
$\sqrt n(\sqrt{\widehat\lambda_R}-c)\Rightarrow
N(0,L_R/(4c^2))$.
To obtain \eqref{eq:rollvar}, note that
\[
\E(\widehat\lambda_R-c^2)^4=O(n^{-2}),\qquad
|\widehat c_R-c|\leq\frac{|\widehat\lambda_R-c^2|}{c}.
\]
Thus $\E\{\sqrt n(\widehat c_R-c)\}^4=O(1)$. The first two moments
of $\sqrt n(\widehat c_R-c)$ converge to those of its normal limit,
which gives \eqref{eq:rollvar}.

For $c_n=hn^{-1/4}$, the same covariance calculation gives
$L_{R,n}\to V^2$. The centered products remain two dependent with
uniformly bounded fourth moments, because $c_n$ is bounded and the
innovations have finite eighth moments. The triangular array result
in Section~\ref{app:probability} and $\sqrt n\,c_n^2=h^2$ imply
\[
\begin{aligned}
\sqrt n\,\widehat\lambda_R&\Rightarrow h^2+VZ,
\qquad Z\sim N(0,1),\\
n^{1/4}\sqrt{\max(0,\widehat\lambda_R)}
&=\sqrt{\max(0,\sqrt n\,\widehat\lambda_R)}
\Rightarrow\sqrt{\max(0,h^2+VZ)}.
\end{aligned}
\]
The last step uses continuity of $v\mapsto\sqrt{\max(0,v)}$ and proves
\eqref{eq:rollboundary}.
\end{proof}

\subsection{Proof of Proposition~\ref{prop:comparison}}\label{app:comparison}

\begin{proposition}\label{prop:comparison}
Fix $c,V>0$ and let $\rho=V/c^2$. Then
\begin{equation}\label{eq:cfvarlimit}
\lim_{a\downarrow0}\frac{\sigma_{\CF}^2(a;c,V)}{c^2}
=1+\frac\rho2+\frac{9\rho^2}{8}
+\frac{3\rho^3}{4}+\frac{3\rho^4}{32}.
\end{equation}
For Roll, as $n \to \infty$ we have
\begin{equation}\label{eq:rollnormalized}
\begin{aligned}
&\frac n{c^2}
\Var\!\left\{\sqrt{\max(0,\widehat\lambda_R)}\right\}\to \frac{5+4\rho+\rho^2}{4}.
\end{aligned}
\end{equation}
The ratio of these coefficients is asymptotic to $3\rho^2/8$ as $\rho\to\infty$.
\end{proposition}

\begin{proof}
Equations~\eqref{eq:app:smallcoefficients} and~\eqref{eq:app:Wvalue} give \eqref{eq:cfvarlimit}. Multiplying \eqref{eq:rollvar} by $n/c^2$ and taking $n\to\infty$ gives $(5+4\rho+\rho^2)/4$. The ratio of the two normalized coefficients is
\[
 \frac{1+\rho/2+9\rho^2/8+3\rho^3/4+3\rho^4/32}
 {(5+4\rho+\rho^2)/4}
 \sim\frac38\rho^2\to\infty.
\]
Hence the CF estimator has no uniform variance advantage.
\end{proof}

\subsection{Small frequency expansions}\label{app:smallfrequency}
Fix $c>0,V\geq0$ and write $m_j=\E(Y_t^j)$, $J_j=Y_t^j-m_j$, and $\rho=V/c^2$. The sign difference is $0,2,-2$ with probabilities $1/2,1/4,1/4$. Gaussian moments therefore give
\begin{equation}\label{eq:app:marginalmoments}
\begin{aligned}
 m_2/c^2&=2+\rho,\\
 m_4/c^4&=8+12\rho+3\rho^2,\\
 m_6/c^6&=32+120\rho+90\rho^2+15\rho^3,\\
 m_8/c^8&=128+896\rho+1680\rho^2\qquad+840\rho^3+105\rho^4.
\end{aligned}
\end{equation}
For the map $g$ in Section~\ref{app:positive}, its partial derivatives at $(x,y)$ satisfy
\[
 g_x=2+m_2a^2+O(a^4),\qquad
 g_y=-\tfrac12-m_2a^2+O(a^4).
\]
For $k=1,2$, the finite twelfth moment controls the remainder in
\[
 \cos(kaY_t)-\phi(ka)
 =-\tfrac12 k^2a^2J_2+\tfrac1{24}k^4a^4J_4+O_{L^2}(a^6).
\]
The order $a^2$ terms cancel in the linear term:
\begin{equation}\label{eq:app:momentinfluence}
\begin{aligned}
 L_a&=g_x\{\cos(aY_t)-x\}+g_y\{\cos(2aY_t)-y\}=a^4Z_c+O_{L^2}(a^6),
 Z_c&=\tfrac32m_2J_2-\tfrac14J_4.
\end{aligned}
\end{equation}
By Lemma~\ref{lem:covariance}, $K=\E(L_a^2)$. Also,
$(S_{11},S_{12},S_{22})=a^4(m_4-m_2^2)(1/4,1,4)+O(a^6)$.
Substitution in \eqref{eq:BK} yields
\begin{equation}\label{eq:app:BKsmall}
\begin{aligned}
 K&=a^8W_c+O(a^{10}),\qquad W_c=\Var(Z_c),\\
 B&=\tfrac34a^4(m_4-m_2^2)+O(a^6).
\end{aligned}
\end{equation}
For the map $h_a$ in Section~\ref{app:positive}, $d=\tan^4(ac)$ gives
\[
 h_a'(d)\sim(4a^4c^3)^{-1},\qquad
 h_a''(d)\sim-3(16a^8c^7)^{-1}.
\]
Using $\sigma_{\CF}^2=h_a'(d)^2K$ and $b=h_a'(d)B+h_a''(d)K/2$ gives
\begin{equation}\label{eq:app:smallcoefficients}
\begin{aligned}
 \lim_{a\downarrow0}\sigma_{\CF}^2&=\frac{W_c}{16c^6},\\
 \lim_{a\downarrow0}b&=\frac{3(m_4-m_2^2)}{16c^3}
                         -\frac{3W_c}{32c^7}.
\end{aligned}
\end{equation}
Finally,
\[
 W_c=\tfrac94m_2^2(m_4-m_2^2)-\tfrac34m_2(m_6-m_2m_4)
       +\tfrac1{16}(m_8-m_4^2),
\]
so \eqref{eq:app:marginalmoments} gives
\begin{equation}\label{eq:app:Wvalue}
\begin{aligned}
 (m_4-m_2^2)/c^4&=4+8\rho+2\rho^2,\\
 W_c/c^8&=16+8\rho+18\rho^2+12\rho^3+\tfrac32\rho^4.
\end{aligned}
\end{equation}
These are fixed parameter limits. No joint limit with sample size is asserted.

\subsection{Proof of Proposition~\ref{prop:rawbias}}\label{app:rawbias}

\begin{proposition}\label{prop:rawbias}
For $c,V>0$,
\begin{equation}\label{eq:rollbiasweighted}
\E\widehat c_R-c=\frac{B_R}{n}+O(n^{-2}),\qquad
B_R=-\frac{L_R}{8c^3}.
\end{equation}
Let $b_0(c,V)=\lim_{a\downarrow0}b(a;c,V)$. Then
\begin{equation}\label{eq:rawbiasregion}
\begin{aligned}
|b_0(c,V)|<|B_R|\iff P(\rho)<0,\\
P(\rho)=9\rho^4+72\rho^3+76\rho^2-80\rho+8.
\end{aligned}
\end{equation}
For $\rho\geq1$, $P(\rho)>0$. Moreover,
\[
\frac{|b_0(c,V)|}{|B_R|}\sim\frac{9\rho^2}{8}
\qquad(\rho\to\infty).
\]
\end{proposition}

\begin{proof}
Let $\Delta_n=\widehat\lambda_R-c^2$. The calculation in
Appendix~\ref{app:roll} and finite dependence give
\[
\E\Delta_n=0,\quad
\E\Delta_n^2=L_R/n+O(n^{-2}),\quad
\E\Delta_n^3=O(n^{-2}),\quad
\E\Delta_n^4=O(n^{-2}).
\]
Gaussian moments and the Bernstein argument in
Appendix~\ref{app:pilotconcentration} give exponentially small tails outside
$|\Delta_n|<c^2/2$. Taylor expansion of the square root through degree three,
with a fourth order remainder, yields
\[
\E\sqrt{\max(0,c^2+\Delta_n)}-c
=-\frac{\E\Delta_n^2}{8c^3}+O(n^{-2})
=-\frac{L_R}{8c^3n}+O(n^{-2}).
\]
Since $b_0<0$ by \eqref{eq:smallbias},
\[
\frac{|b_0|-|B_R|}{c}
=\frac{9\rho^4+72\rho^3+76\rho^2-80\rho+8}{64}.
\]
For $\rho\geq1$ the polynomial is positive: writing $\rho=1+v$ gives
$85+324v+346v^2+108v^3+9v^4$. Its leading term and
$|B_R|/c\sim\rho^2/8$ give the stated ratio.
\end{proof}

\subsection{Proof of Theorem~\ref{thm:gmmlowsnr}}\label{app:gmmlowsnr}

\begin{theorem}\label{thm:gmmlowsnr}
Fix $V>0$ and distinct $x_1,\ldots,x_J>0$, $J\geq2$. Choose
$u_j=x_j/\sqrt V$ and hold these frequencies fixed during minimization.
For each sufficiently small $c>0$, use $W=\Omega(c,V)^{-1}$, held fixed, and a bounded
extension of the local solution outside its smooth domain. Define
\[
\begin{aligned}
B_G(c,V)&=\lim_{n\to\infty}n(\E\widehat c_G-c),\\
\sigma_G^2(c,V)&=\lim_{n\to\infty}n\Var(\widehat c_G),
\end{aligned}
\]
where $\widehat c_G$ is the marginal ECF--GMM spread estimate in
Section~\ref{sec:gmm}.
Put
\[
\begin{aligned}
A_{jk}&=\cosh(x_jx_k)-1,\\
v_j&=-x_j^2/2,\qquad g_j=-x_j^4/3,\\
I_0&=g^{\mathsf T}A^{-1}g
-\frac{(g^{\mathsf T}A^{-1}v)^2}{v^{\mathsf T}A^{-1}v}.
\end{aligned}
\]
Then $0<I_0\leq8/3$ and, as $c\downarrow0$,
\begin{equation}\label{eq:gmmlowcoefficients}
B_G\sim-\frac{3V^4}{8I_0c^7},\qquad
\sigma_G^2\sim\frac{V^4}{4I_0c^6}.
\end{equation}
Consequently,
\begin{equation}\label{eq:gmmlowratios}
\frac{|B_G|}{|B_R|}\sim\frac{3\rho^2}{I_0},\qquad
\frac{\sigma_G^2}{L_R/(4c^2)}\sim\frac{\rho^2}{I_0}.
\end{equation}
For $J=2$, $u_2=2u_1$, and $0<u_1c<\pi/4$, the local spread solution
is \eqref{eq:est}.
\end{theorem}

\begin{proof}
Use parameters $T=V+2c^2$ and $q=c^4$. Uniformly over the fixed finite
frequency set,
\begin{equation}\label{eq:gmmlocalexpansion}
f_j(T,q)=e^{-Tu_j^2/2}\left\{1-\frac{u_j^4q}{6}
-\frac{2u_j^6q^{3/2}}{45}+O(q^2)\right\}.
\end{equation}
At $q=0$, let $P=\operatorname{diag}(e^{-x_j^2/2})$. Then
\[
\Omega\longrightarrow PAP,\qquad
\partial_Tf\longrightarrow Pv/V,\qquad
\partial_qf\longrightarrow Pg/(2V^2).
\]
The matrix $A$ is positive definite because distinct cosine functions are
linearly independent under a Gaussian law. The vectors $v,g$ are linearly
independent for distinct positive $x_j$. Write $B_q$ and $\sigma_q^2$
for the leading bias and variance coefficients of the fitted $q$ coordinate.
Taking a Schur complement gives
\begin{equation}\label{eq:gmmqvariance}
\sigma_q^2\longrightarrow\frac{4V^4}{I_0},\qquad I_0>0.
\end{equation}

For $Z\sim N(0,1)$, put
$K_j=e^{x_j^2/2}\cos(x_jZ)-1$ and use the Hermite polynomials
$H_2(Z)=Z^2-1$, $H_4(Z)=Z^4-6Z^2+3$. Then
\[
\E(KK^{\mathsf T})=A,\quad
\E(KH_2/2)=v,\quad
\E(-KH_4/3)=g,\quad
\E(H_4^2)/9=8/3.
\]
The squared norm of the projection of $-H_4/3$ onto the cosine span is
$g^{\mathsf T}A^{-1}g\leq8/3$. Removing its nuisance direction subtracts
the nonnegative term defining $I_0$, hence $I_0\leq8/3$.

For $p=(T,q)$, write $D_p=\partial f/\partial p^{\mathsf T}$ and
\[
\mathcal I=D_p^{\mathsf T}\Omega^{-1}D_p,\qquad
L=\mathcal I^{-1}D_p^{\mathsf T}\Omega^{-1},\qquad
h_j=\operatorname{tr}\{\nabla_p^2f_j\mathcal I^{-1}\}.
\]
For a centered moment error with covariance $\Omega$,
$L\Omega(I-D_pL)^{\mathsf T}=0$. A second order expansion of the normal
equations consequently gives the bias vector
\[
\boldsymbol B_p=-\tfrac12\mathcal I^{-1}D_p^{\mathsf T}\Omega^{-1}h.
\]
The limiting information is positive definite, and \eqref{eq:gmmlocalexpansion} implies
$\|\nabla_p^2f_j\|=O(q^{-1/2})$. Hence $B_q=O(q^{-1/2})$.
The transformation $c=q^{1/4}$ gives
\[
B_G=\frac{B_q}{4q^{3/4}}-\frac{3\sigma_q^2}{32q^{7/4}}
=O(c^{-5})-\frac{3\sigma_q^2}{32c^7},\qquad
\sigma_G^2=\frac{\sigma_q^2}{16c^6}.
\]
Equations~\eqref{eq:gmmqvariance} and \eqref{eq:rollbiasweighted} prove
\eqref{eq:gmmlowcoefficients} and \eqref{eq:gmmlowratios}.
With two frequencies $u,2u$, the two moment equations are locally invertible.
Their spread solution is \eqref{eq:est}, independently of $W$.
\end{proof}

\end{document}